\documentclass[%
reprint,
superscriptaddress,
amsmath,amssymb,
aps,
floatfix,
]{revtex4-2}
\usepackage{graphicx}
\usepackage[caption=false]{subfig}
\usepackage{siunitx}
\usepackage{csquotes}
\usepackage{bm}
\usepackage{physics}
\usepackage{mathtools}
\usepackage{amsfonts}
\usepackage{amssymb}
\usepackage{amsmath}
\usepackage{diagbox}
\usepackage{hyperref}
\usepackage{xspace}
\hypersetup{linktocpage,colorlinks,citecolor={blue},pdfdisplaydoctitle=true,pdfpagemode=UseOutlines,bookmarksnumbered=true}
\usepackage{dsfont}
\usepackage{todonotes}
\usepackage{booktabs}
\usepackage[most]{tcolorbox}

\newcommand{\clbound}{\ensuremath{\beta_{C}}\xspace}
\newcommand{\epscrit}{\ensuremath{\epsilon^{*}}\xspace}
\newcommand{\qbound}{\ensuremath{\beta_{Q}}\xspace}
\newcommand{\LDSset}{\mathcal{S}_{\text{LDS}}}

\begin{document}
\title{Effects of Topological Boundary Conditions on Bell Nonlocality}
\author{Patrick Emonts}
\thanks{These two authors contributed equally}
\affiliation{Instituut-Lorentz, Universiteit Leiden, P.O. Box 9506, 2300 RA Leiden, The Netherlands}
\author{Mengyao Hu}
\thanks{These two authors contributed equally}
\affiliation{Instituut-Lorentz, Universiteit Leiden, P.O. Box 9506, 2300 RA Leiden, The Netherlands}
\author{Albert Aloy}
\affiliation{ Institute for Quantum Optics and Quantum Information, Austrian Academy of Sciences, Boltzmanngasse 3, A-1090 Vienna, Austria}
\affiliation{Vienna Center for Quantum Science and Technology (VCQ), Faculty of Physics, University of Vienna, Vienna, Austria}
\author{Jordi Tura}
\affiliation{Instituut-Lorentz, Universiteit Leiden, P.O. Box 9506, 2300 RA Leiden, The Netherlands}
\date{\today}

\begin{abstract}
    Bell nonlocality is the resource that enables device-independent quantum information processing tasks. 
    It is revealed through the violation of so-called Bell inequalities, indicating that the observed correlations cannot be reproduced by any local hidden variable model. 
    While well explored in few-body settings, the question of which Bell inequalities are best suited for a given task remains quite open in the many-body scenario.
    One natural approach is to assign Bell inequalities to physical Hamiltonians, mapping their interaction graph to two-body, nearest-neighbor terms.
    Here, we investigate the effect of boundary conditions in a two-dimensional square lattice, which can induce different topologies in lattice systems.
    We find a relation between the induced topology and the Bell inequality's effectiveness in revealing nonlocal correlations.
    By using a combination of tropical algebra and tensor networks, we quantify their detection capacity for nonlocality.
    Our work can act as a guide to certify Bell nonlocality in many-qubit devices by choosing a suitable Hamiltonian and measuring its ground state energy; a task that many quantum experiments are purposely built for.
\end{abstract}

\maketitle

\section{Introduction\label{sed:introduction}}

Nonlocality is a fundamental characteristic of Nature in which the statistics obtained by taking certain local measurements on some composite (quantum) systems cannot be replicated by any local hidden variable model~\cite{bell_einstein_1964}.
No local deterministic strategy, even if aided by shared randomness, can replicate these so-called nonlocal correlations~\cite{fine_hidden_1982}.
The violation of a Bell inequality witnesses the presence of nonlocality~\cite{brunner_bell_2014}, and it has recently been proven in several loophole-free Bell experiments~\cite{hensen_loophole-free_2015, giustina_significant-loophole-free_2015, shalm_strong_2015, rosenfeld_event-ready_2017, storz_loophole-free_2023}.  
Nonlocal correlations allow to detect entanglement from a minimal set of assumptions, under the so-called device-independent (DI) paradigm.
Nonlocality is the resource underpinning the implementation of DI quantum information protocols~(DIQIP) such as DI quantum key distribution~\cite{acin_device-independent_2007, pironio_security_2013}, DI randomness amplification~\cite{colbeck_free_2012, gallego_full_2013}, or DI self-testing~\cite{supic_self-testing_2016}. 
In the multipartite scenario, however, a thorough understanding of the emergence of nonlocal correlations remains elusive.

Motivated by the advent of DIQIP, designing operationally useful Bell inequalities has been a topic of intensive research in the last years. 
To distinguish classical from nonlocal correlations, it is convenient to study correlations in terms of local hidden-variable models (LHVMs).
The correlations of LHVMs can be characterized by a polytope. 
The characterization of the LHVM polytope gives a geometrically complete and minimal description of the LHVM set in terms of the so-called facets~\cite{rosset_classifying_2014}, or tight Bell inequalities.
However, these quickly lose the properties that make them appealing for DIQIP tasks, even as the parameters of the Bell scenario only slightly increase. 
For instance, the CHSH inequality~\cite{clauser_proposed_1969} is tight and self-tests the maximally entangled state of two qubits, but its facet generalization to more outcomes, such as the CGLMP inequality~\cite{collins_bell_2002}, loses that property and needs to be tilted~\cite{masanes_tight_2003}. 
Therefore, even the computationally prohibitive method of finding all facet Bell inequalities is no guarantee to yield a useful result. 
Roughly speaking, the underlying reason may be there is nothing quantum in the definition of a LHV model.

In the multipartite case, the complexity of finding all facets would scale as $O(\exp(\exp(n)))$ for $n$ parties, so radically different techniques are necessary~\cite{chazelle_optimal_1993}.
Restricting the study to those Bell inequalities that consist of few-body correlators and inherit the geometry of the problem is a good trade-off between complexity and representability, further allowing for the experimental implementation of Bell correlation witnesses~\cite{baccari_efficient_2017, schmied_bell_2016, engelsen_bell_2017}. 
There, one can establish a natural connection to local Hamiltonians, looking at their ground state energy to witness nonlocality. 
However, this is not a one-to-one correspondence: a Bell inequality may be associated to many Hamiltonians, and a Hamiltonian may act as a particular Bell operator of many inequalities.
Here, a Bell operator corresponds to the quantum operator resulting from a choice of measurements for a given Bell inequality.

The question of optimal constituents of a good Bell inequality for multipartite correlations is rather open-ended, typically involving multiple optimizations. 
We know that a certain degree of frustration must be present among the correlators: e.g., the minus sign in the CHSH inequality $\langle A_0B_0 \rangle + \langle A_0B_1 \rangle + \langle A_1B_0 \rangle - \langle A_1B_1 \rangle \leq \beta$ guarantees that the LHVM (classical) bound, $\beta_{\text{c}} = 2$, is strictly smaller than the algebraic bound, $\beta_{\text{alg}}=4$~\cite{brunner_bell_2014}. 
Otherwise, we get a completely trivial inequality. 
As a straightforward generalization, one might consider a multi-partite system on a one-dimensional lattice. 
On every site, one might replicate a $O(1)$-partite inequality, such as the CHSH, with e.g. alternating weights depending on some coupling parameter, thus creating a multipartite inequality~\cite{tura_energy_2017}, which can be seen as a dimer covering of the 1D lattice.
In contrast, with an homogeneous choice, one replicates the case of translationally invariant inequalities with $O(1)$-nearest-neighbor interactions, a method that has allowed to characterize infinite, translationally invariant inequalities in 1D for some scenarios~\cite{wang_entanglement_2017}.

In one spatial dimension the number of possibilities is rather limited: one can only place dimers in the e.g. even-odd links and only has the choice of open vs. periodic boundary conditions and desirable properties due to finite-size effects wash out in the thermodynamic limit~\cite{tura_translationally_2014}.

In 2D lattices, the number of possibilities vastly increases, but so does the computational complexity. 
The number of dimer coverings increases exponentially with the number of sites and the possibilities for boundary conditions also multiply, yielding different underlying topologies, which can be classified through the fundamental theorem of compact surfaces. 
Hence, one should expect that the particular arrangement of the dimer covering may affect the robustness of the resulting multipartite inequality. 
We show that this is indeed the case and identify qualitatively and quantitatively in which way this happens for $3\times 3$, $4\times 4$ and $5\times 5$ square lattices and boundary conditions corresponding to a torus and a Klein bottle.

In this work, we propose a method to gain insights into the interplay between frustration, boundary conditions and finite-size effects for nonlocality detection.
We therefore focus our study on square 2D lattices. 
Due to the immense number of free parameters in such a problem, we restrict ourselves to a CHSH inequality in every link, with mutually unbiased measurements at every site.
We build the multipartite Bell inequality through a dimerization procedure: By picking a dimer covering of the lattice, we place a higher weight on the links with a dimer. 
The resulting Bell operator can be straightforwardly interpreted as a local Hamiltonian.
We compute the classical bound and quantum value of the Bell operator with a combination of tropical algebra and tensor network methods~\cite{hu_tropical_2022}.

We find that the sole arrangement of dimers has a pronounced effect on the capacity of the Bell inequality to detect nonlocal correlations.
We compute the detection capacity of individual inequalities by comparing their classical bound and their quantum value.
The detection capacity for small systems shows clear differences depending on the boundary conditions of the system.
By analyzing the robustness of the inequality to noise on the coupling parameter, we suggest the most robust configurations for on-device testing.

The rest of this paper is organized as follows:
In Sections~\ref{sec:bell_inequalities} and~\ref{sec:connection_ham} we give an introduction to Bell inequalities and how they are connected to Hamiltonians.
Section~\ref{sec:methods} explains the methods to obtain and analyze the two-dimensional models.
The results are presented in Section~\ref{sec:results}.
Finally, we conclude and present an outlook in Section~\ref{sec:conclusion}.

\section{Preliminaries --- Bell inequalities\label{sec:bell_inequalities}}
The traditional scenario to observe nonlocality are two distant parties, Alice and Bob, who can perform measurements on a shared physical system.
Each of them chooses among $m$ different measurements and each measurement can yield $d$ possible outcomes.
Here, we will directly start with $N$ parties sharing a physical system since we are interested in many-body nonlocality.
We denote the measurement choices of all parties with $x_i \in [m]= \{0, \dots , m-1\}$ and their respective outputs $a_i \in [d]=\{0,\dots ,d-1\}$ for $i\in [N]=\{0,\dots,N-1\}$, respectively.
In general, the correlations between results obtained in the above process are governed by the conditional probability distribution $P(\vb{a}\vert \vb{x}):=P(a_0,\dots ,a_{N-1}\vert x_0,\dots ,x_{N-1})$ with $\vb{a}:=a_0,\dots,a_{N-1}$ and $\vb{x}:=x_0,\dots,x_{N-1}$. 
This joint probability distribution is fully described by the $(md)^N$ dimensional vector  
\begin{align}
    \{P(a_0,\dots ,a_{N-1}\vert x_0,\dots ,x_{N-1})\}_{\vb{a};\vb{x}}.
    \label{eq:jointprob}
\end{align}
All entries satisfy the $m^N$ affine-linear equations
\begin{align}
    \begin{split}
        \label{eq:probsSumToOne}
    \sum_{\vb{a}}P(a_0,\dots ,a_{N-1}\vert x_0,\dots ,x_{N-1}) = 1, \; \forall \vb{x},
    \end{split}
\end{align}
to be normalized as well as the inequalities $P(\vb{a} \vert \vb{x}) \geq 0$ to be non-negative.

The non-signaling principle~\cite{cirelson_quantum_1980,popescu_quantum_1994} (all parties are spatially separated and cannot communicate instantaneously), leads to well-defined marginals, i.e. the marginals observed by any subset of parties do not depend on the choices of measurements performed by the rest. 
That is, for all $x_i,x'_i$,
\begin{align}
    \begin{aligned}
        &\sum_{a_i}P(a_0,\dots ,a_i,\dots ,a_{N-1}\vert x_0,\dots ,x_i,\dots ,x_{N-1}) \\
        &= \sum_{a_i}P(a_0,\dots ,a_{i},\dots ,a_{N-1}\vert x_0,\dots ,x'_i,\dots ,x_{N-1}).
    \end{aligned}
\end{align}
Thus, $P(a_{i_1},\dots ,a_{i_l}\vert x_{i_1},\dots ,x_{i_l})$ is well defined on any subset $\{i_1,\dots ,i_l\}\subseteq [N]$.

To detect nonlocality, the goal is to find Bell inequalities separating the polytope $\mathcal{L}$ of local correlations described by LHVM from the convex set of quantum correlations. 
LHVM can be formulated as 
\begin{align}
        \label{eq:p_local_corr}
    P(\vb{a} \vert \vb{x})=\sum_{\lambda}p(\lambda)\prod_{i=0}^{N-1}P(a_i | x_i,\lambda),
\end{align}
where $\lambda$ is some hidden variable. 
Among LHVM correlations, of special interest are local deterministic strategies (LDS)~\cite{fine_hidden_1982}, which correspond to the vertices of the LHVM polytope and factorize as
\begin{align}
    P(a_0,\dots ,a_{N-1}\vert x_0,\dots ,x_{N-1})=\prod_{i=0}^{N-1}P(a_i\vert x_i),
    \label{eq:LHV}
\end{align}
where $P(a_i\vert x_i)$ are deterministic functions.

In contrast, the quantum correlations are given by Born's rule:
\begin{equation}
    \begin{split}
         \label{eq:quantumcorrelations}
    &P(a_0,\dots ,a_{N-1}\vert x_0,\dots ,x_{N-1
    })\\
    =&\text{Tr}[\rho_N(\mathcal{M}^{a_0}_{0,x_0}\otimes \dots  \otimes \mathcal{M}^{a_{N-1}}_{N-1,x_{N-1}})],
    \end{split}
\end{equation}
where $\rho_N$ is an $N$-partite quantum state and $\mathcal{M}_{i,x_i}^{a_i}\succeq 0$ is the $x_i$-th measurement with outcome $a_i$ performed by party $A_i$, satisfying the normalization condition $\sum_{a_i}\mathcal{M}_{i,x_i}^{a_i}=\mathbb{I}$. 
The convex set of such quantum correlations is denoted by $\mathcal{Q}$.

Tight Bell inequalities are facets of the local polytope $\mathcal{L}$ separating local correlations from quantum correlations.
These inequalities take the form $\mathcal{I}_{N,m,d} \geq \clbound$, where
\begin{align}
    \label{eq:BellNmd}
    \mathcal{I}_{N,m,d}:=\sum_{\vb{a},\vb{x}}\alpha_{\vb{a},\vb{x}}P(\vb{a} \vert \vb{x}) ,
\end{align}
and $\alpha_{\vb{a},\vb{x}} \in \mathbb{R}$ are some coefficients and $\clbound=\min_{P(\vb{a}\vert \vb{x}) \in \mathcal{L}} I_{N,m,d}$ is the classical bound of the Bell inequality. 
Note that, depending on the underlying physical model, a given set of correlations might or might not be compatible with it. 
For instance, if the model is LHVM, $\mathcal{I}_{N,m,d}$ is lower-bounded by $\beta_C$. 
Similarly, if we consider quantum theory as the underlying physical model, the inequality $\mathcal{I}_{N,m,d} \geq \qbound$, where $\qbound = \inf_{P(\vb{a}\vert \vb{x}) \in \mathcal{Q}} I_{N,m,d}$, probes the limits of the quantum set $\mathcal{Q}$. 
It is often referred to as the quantum Bell inequality, with $\qbound$ representing the associated quantum bound or Tsirelson bound~\cite{cirelson_quantum_1980}.

Naturally, it is desirable to have a maximal distance between \clbound and \qbound to facilitate the detection of Bell nonlocality in experiments.
We can compute the classical bound of $\mathcal{I}_{N,m,d}$ by minimizing it over all the LDSs of $N$ parties.
An LDS of a single party can be described by a map $\phi_s:[m]\mapsto [d]$ that deterministically associates an outcome to every input.
We can enumerate all $d^m$ possible strategies as the set of LDS $\LDSset
:=[d^m]$. 
In the case of the CHSH inequality with $m=2$ measurements and $d=2$ (binary) outcomes, the set of LDS is $\LDSset = \{0,\dots,3\}$. 
The input-output maps are $\phi_0(x)=0$,  $\phi_1(x)=x$, $\phi_2(x)=1-x$ and $\phi_3(x)=1$.
In words, the strategies are equivalent to choosing always $0$, choosing the input, choosing the opposite of the input or always choosing $1$, respectively.
Now the probability distribution of a single party associated with the LDS $s\in \LDSset$ is given by the Kronecker delta $P_{s}(a \vert x):= \delta(a-\phi_{s}(x))$.
For an $N$-partite strategy vector $\boldsymbol{s} \in \LDSset^N$, the joint probability distribution of LDS can then be written as
$P_{\boldsymbol{s}}(a_0,\dots,a_{N-1} \vert x_0,\dots,x_{N-1}) := \prod_{i=0}^{N-1} P_{s_i}(a_i|x_i)$. 
Thus Eq.~\eqref{eq:BellNmd} can be rephrased as 
\begin{align}
    \label{eq:lds_Bell_ineq}
    \mathcal{I}_{N,m,d}(\vb{s}):=\sum_{\vb{a},\vb{x}}\alpha_{\vb{a},\vb{x}}P_{\vb{s}}(\vb{a} \vert \vb{x}),
\end{align}
where $\vb{s} \in \LDSset^N$ is an $N$-partite strategy vector, and the classical bound 
\begin{align}
    \label{eq:clbound_lds}
    \clbound=\min_{P(\vb{a}\vert \vb{x}) \in \mathcal{L}} I_{N,m,d}=\min_{\vb{s}\in \LDSset^N} \mathcal{I}_{N,m,d}(\vb{s}).
\end{align}
We will use this expression to compute the classical bound in Section ~\ref{sec:classical_bound}.

\section{The Model\label{sec:connection_ham}}
The relationship between Bell inequalities and Hamiltonians provides a way for testing nonlocality and designing quantum systems to demonstrate nonlocal correlations~\cite{tura_detecting_2014,tura_nonlocality_2015,tura_energy_2017, baccari_efficient_2017,wang_entanglement_2017,fadel_bell_2018,fadel_bounding_2017,wagner_bell_2017,wang_two-dimensional_2018}. 
For a given Bell inequality, one can construct a Hamiltonian coinciding with its Bell operator, which provides a way to optimize the many-body system to exhibit nonlocal correlations.
If the ground state energy of the Hamiltonian is smaller than the classical bound given by the Bell inequality, it certifies that the quantum system exhibits non-local correlations.
Conversely, we can also associate a Bell inequality to a given spin Hamiltonian such that its Bell operator coincides with it. 
This correspondence is not one-to-one, for details see Appendix~\ref{app:chained_bell}.

\begin{figure}
    \includegraphics[width=\columnwidth]{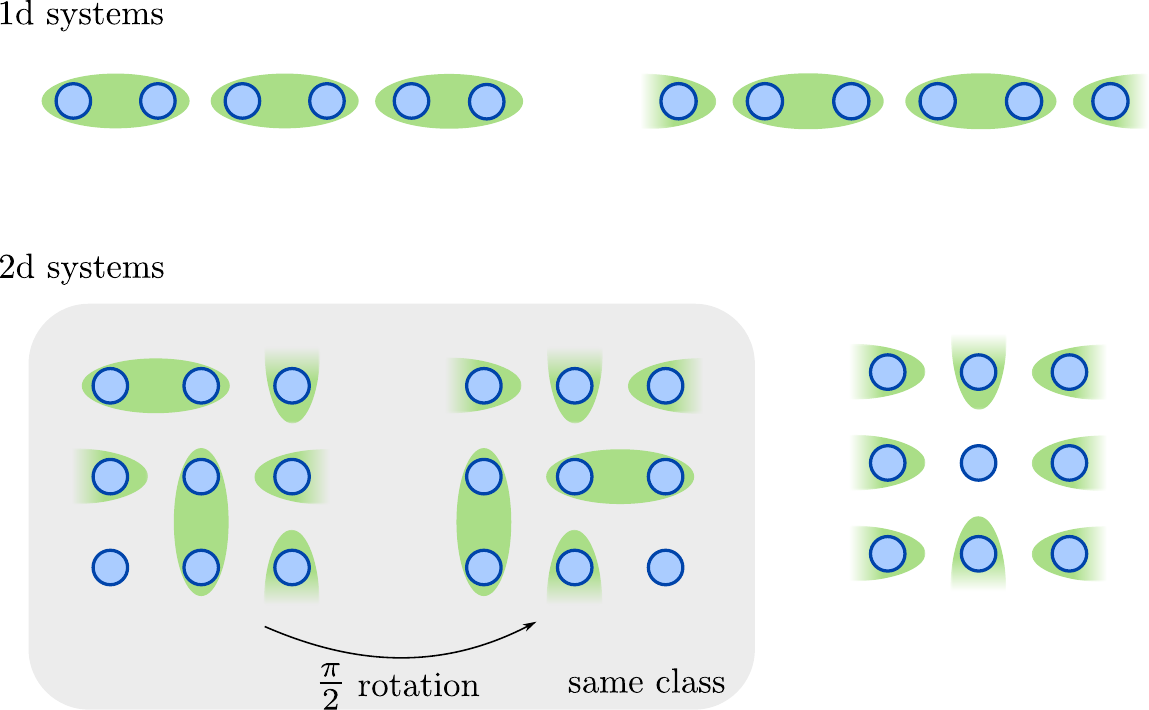}
    \caption{Dimer configurations in one and two dimensions with periodic boundary conditions.
        In the one-dimensional case, there are only two distinct dimer coverings (links on even or odd links).
        In two spatial dimensions, multiple dimer configurations are possible.
        Dimer configurations related by spatial symmetries (see grey box) yield the quantum and classical bounds.
    }
    \label{fig:dimer_configurations}
\end{figure}

One example for a many-body inequality is the CHSH Bell expression realized on every link of a $1$D system.
On a chain with $N$ qubits ($N$ even), one can construct a $1$D Bell inequality $I^{(N)}(\epsilon)=\sum_{i=0}^{N-1}f_i(\epsilon)\mathcal{I}^{(i,i+1)}$, where $\mathcal{I}^{(i,i+1)}$ is the CHSH inequality between sites $i$ and $i+1$ and $f_i(\epsilon)$ are real coefficients of the multi-partite inequality.
The Hamiltonian 
\begin{equation}
    \begin{split}
 \label{eq:1D_hamiltonian}
        H=\sum_{i=0}^{N-1}f_i(\epsilon)&(\sigma_x^{(i)}\sigma_x^{(i+1)}+\sigma_x^{(i)}\sigma_z^{(i+1)}\\&+\sigma_z^{(i)}\sigma_x^{(i+1)} -\sigma_z^{(i)}\sigma_z^{(i+1)})
    \end{split}
\end{equation}
with $f_i(\epsilon)=1+(-1)^i\epsilon$ is a special case of the multipartite Bell inequality $I^{(N)}(\epsilon)$ which we choose in this work.
For ease of description and fair comparison between different systems, we choose a version of the CHSH Bell operator with equal measurements at all sites throughout this work~\cite{tura_energy_2017}, i.e. $A_0=B_0=\sigma_x$ and $A_1=B_1=\sigma_z$.
Thus, Equation~\eqref{eq:1D_hamiltonian} is equivalent to a weighted CHSH expression on each link.
To obtain a Bell violation with identical measurements for both parties, we use a rotated Bell state $\ket{\phi}=\frac{1}{\sqrt{2}}(\ket{00}+e^{\pi\mathbf{i}/4}\ket{11})$.
The structure of the coefficients $f_i(\epsilon)$ follows a dimer pattern: every second link has a strong coupling, the others are only weakly coupled (cf. Fig.~\ref{fig:dimer_configurations}).
This construction has two interesting limiting cases: $\epsilon=0$ and $\epsilon=1$.
For $\epsilon=0$, the multipartite Bell inequality $I (\epsilon)$ corresponds to a sum of CHSH inequalities with the same weights between all the neighboring nodes: $I(0)=\sum_{i=0}^{N-1} \mathcal{I}^{(i,i+1)}$. 
The inequality $I(0)$ cannot be violated due to monogamy of Bell correlations~\cite{toner_monogamy_2006,cieslinski_unmasking_2024}, which implies that the violation of CHSH inequality between nodes $i$ and $i+1$ prohibits any violation of other CHSH inequalities involving nodes connected to nodes $i$ and $i+1$. 

In the other limiting case $\epsilon=1$, the Bell inequality $I (\epsilon)$ corresponds to a sum of disjoint CHSH inequalities on all links emanating from an even node.
The expression $I (1)=2\sum_{i}\mathcal{I}^{(i,i+1)}$ is maximally violated by the state $\bigotimes_{i}\ket{\phi_2}_{i,i+1}$, where $\ket{\phi_2}=\frac{1}{\sqrt{2}}(\ket{00}+e^{\pi\mathbf{i}/4}\ket{11})$, and the index $i$ only iterates over even sites.
Moreover, the inequality is maximally violated with a quantum value of $-N\sqrt{2}$.

While we know the point of maximal violation at $\epsilon=1$, the range of the couplings $\epsilon$ with a possible violation is not known.
In one dimension, the classical bound $\clbound(\epsilon)$ of $I^{(N)}(\epsilon)$ can be obtained by contracting a tropical tensor network, for details see Section~\ref{sec:classical_bound}.
Finding the quantum value $\qbound(\epsilon)$ can be reduced to finding the ground state energy of the Hamiltonian $H$ in Eq.~\eqref{eq:1D_hamiltonian}. 
Here, we are careful to call the ground state of the Hamiltonian the \emph{quantum value}, since we keep a fixed set of measurements during the minimization.
The \emph{quantum bound} is the infimum over all measurements and states in the quantum set.
The nonlocality is detected when $\qbound(\epsilon)<\clbound(\epsilon)$.

In a single spatial dimension, the choice of inequalities is heavily restricted by the limited choice of boundary conditions and dimer configurations. 
In 1D, there are only two boundary conditions: open or periodic boundary conditions.
Additionally, the weighted links can either be on the odd or on the even links, leading to only four possibilities in total.

The variety of inequalities changes dramatically in two spatial dimensions.
Instead of the two possible dimer coverings as in one dimension, the number of allowed dimer coverings scales exponentially with system size in two dimensions~(cf. Fig.~\ref{fig:dimer_configurations}).
Additionally, different boundary conditions are possible.
According to the classification theorem for compact, connected surfaces, every compact connected surface is homeomorphic to the connected sum of torus, Klein bottle and sphere~\cite{gallier_classification_2013}.
We limit our study to the topologies of torus and Klein bottle, since the sphere is not well-suited for a square lattice.

The inequality of the 1D case can be readily generalized to the two dimensional case.
Instead of a fixed link pattern (stronger weights on the even links) as in the one-dimensional case, we choose a fixed dimer covering on the square lattice and place the weighted links on the dimers. 
Thus one can construct the multipartite Bell expression for a given dimer configuration
\begin{align}
     \label{eq:multiBI2D_construct}
        I(\epsilon):=\sum_{\langle i,j\rangle}f_{i,j} \cdot \mathcal{I}^{(i,j)},
\end{align}
where 
\begin{align}
    \label{eq:fij(epsilon)}
    f_{i,j}(\epsilon)=\begin{cases}
        1+\epsilon \qq{if $(i,j)$ in dimer,}\\
        1-\epsilon \qq{if $(i,j)$ not in dimer,}
    \end{cases}
\end{align}
and $\langle i,j\rangle$ indicates that nodes $i$ and $j$ are nearest neighbours. 
The two limiting cases of this construction are the same as the limiting cases of the 1D construction.
Following the same strategy as in the one-dimensional case, we construct the corresponding Hamiltonian
\begin{equation}
    \begin{split}
        \label{eq:fromHtoBellineq}
        H:=\sum_{\langle i,j\rangle}f_{i,j}(\epsilon)&(\sigma_x^{(i)}\sigma_x^{(j)}+\sigma_x^{(i)}\sigma_z^{(j)}\\&+\sigma_z^{(i)}\sigma_x^{(j)} -\sigma_z^{(i)}\sigma_z^{(j)})
    \end{split}
\end{equation}
where $f_{i,j}(\epsilon)$ is defined as in Eq.~\eqref{eq:fij(epsilon)}. 

\begin{figure}
    \centering
    \includegraphics[width=\columnwidth]{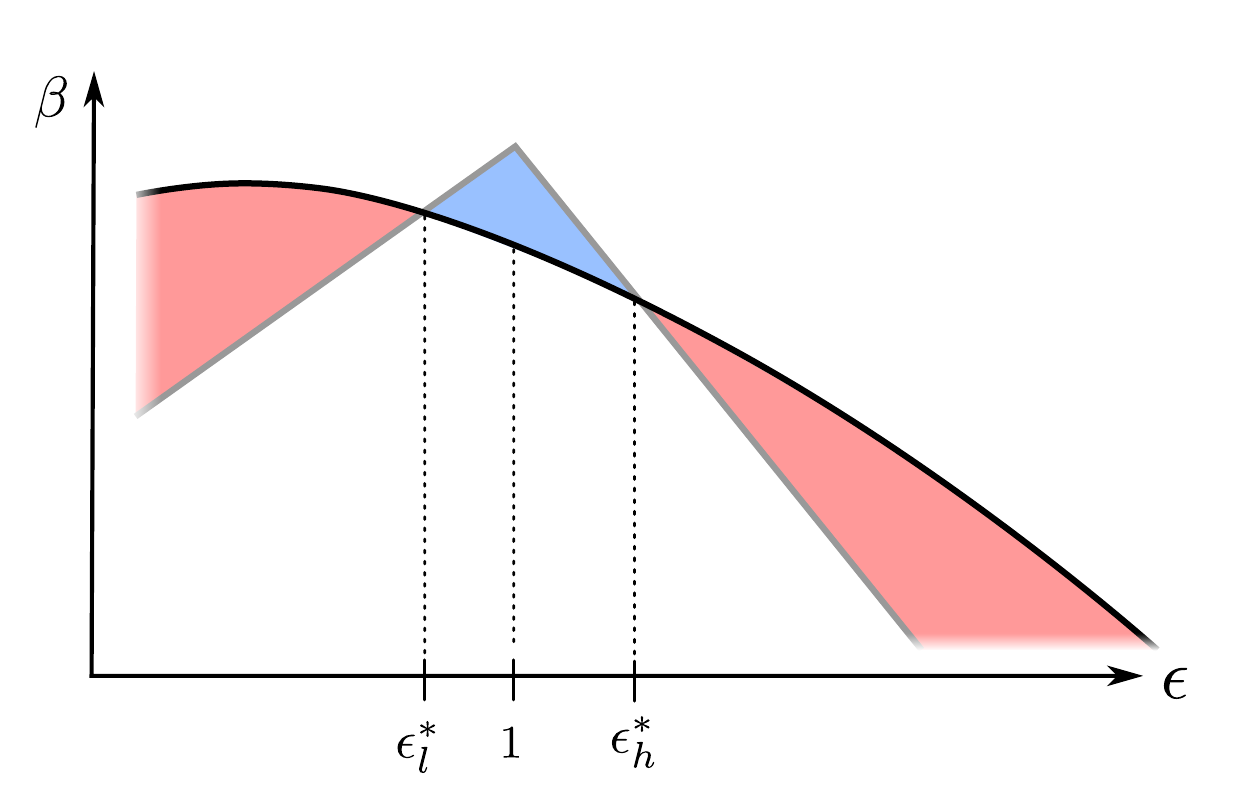}
    \caption{Possible violation of a bell inequality. Depending on $\epsilon$, the classical bound \clbound (gray line) and the quantum value $\qbound$ (black line) of the inequality vary.
    The region where a violation is possible (in blue) depends on the size of the system, on the dimer configuration and the boundary conditions.
    The intersections between the classical bound and the quantum value are the critical values of the coupling $\epscrit_l$ and $\epscrit_h$.
    }
    \label{fig:sketch_epsilon}
\end{figure}

To detect nonlocality, we are again interested in cases where the Bell expression has a quantum value \qbound and a classical bound \clbound such that  $\qbound(\epsilon)<\clbound(\epsilon)$.
As discussed in Ref.~\cite{tura_energy_2017}, the dimer model allows for violations of Bell inequalities in a region around $\epsilon=1$, where $\epsilon=1$ corresponds to one of the limiting cases discussed above.
Figure~\ref{fig:sketch_epsilon} shows the behavior in a sketch.
A violation of the Bell inequality can only be detected inside of the blue-shaded region.
The boundaries of the region in terms of $\epsilon$ are given by the intersections of the quantum value and the classical bound
\begin{align}
    \qbound(\epscrit)/\clbound(\epscrit)=1.
    \label{eq:def_eps_crit}
\end{align}
We denote these values of the coupling $\epscrit$ on either side of $\epsilon=1$ by $\epscrit_l$ and $\epscrit_h$, respectively.
The size and shape of the region with a violation depends on the chosen dimer configuration, the size of the system and the boundary conditions.
The larger the distance between $\epscrit_h$ and $\epscrit_l$, the higher the capacity of the inequality to detect nonlocality.
The goal is to find a combination of dimer configuration and boundary condition where the interval with a violation is maximal.
These configurations are prime candidates for experimental realization to certify nonlocality.

\section{Methods\label{sec:methods}}
To maximize the size of the violating region, we first need a way to compute the two bounds of the violating region $\epscrit_l$ and $\epscrit_h$.
If it does not matter whether we treat the lower or upper bound, we will refer to both of them jointly with $\epscrit$.
The algorithm to estimate the critical values of epsilon proceeds in two major steps.
First, we address the dependence of the dimer configuration.
All dimer coverings can be classified with regard to the symmetries of the lattice. 

A simple example of equivalent dimer coverings are two patterns that are rotated by 90 degrees on a lattice with periodic boundary conditions.
These two coverings will give the same values for \epscrit and we only have to compute them once.
We explore the symmetries of the lattices in~\ref{sec:dimer_classification}.

The second step is the search for the actual value of \epscrit.
Since \epscrit is a quotient of a classical bound and a quantum value, we have to compute both of them.
By iteratively computing the classical (cf.~\ref{sec:classical_bound}) and the quantum value(cf.~\ref{sec:quantum_bound}), we can find the intersection between the two bounds in terms of $\epsilon$, the critical value $\epscrit$ (cf.~\ref{sec:critical_epsilon}).

\subsection{Classification of dimer coverings\label{sec:dimer_classification}}
Let us start with the dimer coverings.
In one dimension, as considered in~\cite{tura_energy_2017}, there are only two different dimer coverings.
The dimers can be on even or on odd links (cf. Figure~\ref{fig:dimer_configurations}).

In two dimensions, the number of possible dimer configurations scales with the size of the lattice.
Since we are interested in maximal violations, we focus on maximal dimer coverings.
A maximal dimer covering distributes as many dimers on a given lattice as is allowed.

However, not all maximal dimer coverings are independent.
If two dimer coverings are connected by a symmetry $U$ that commutes with the Hamiltonian $[U,H]=0$, the quantum value will not be affected.
The unitaries $U$ are generated by lattice symmetries $O$ like rotations or mirroring (cf. Figure~\ref{fig:dimer_configurations}).

As an example, we will consider the symmetries of the torus. 
In total, the torus has five non-decomposable symmetries $O$: right shift, up shift, rotation by 90 degrees, horizontal mirror, vertical mirror.
Details about the symmetry considerations and the symmetries of the Klein bottle are given in Appendix~\ref{app:topology}.
For concreteness, we focus on a specific symmetry, the right shift $O_{rs}$ (\emph{r}ight \emph{s}hift).
The operation right shift is a bijection $O_{rs}$ from the set $\mathbb{S}_{n,T}$ to the set $\mathbb{S}_{n,T}$,
where $\mathbb{S}_{n,T}$ is the set of all maximal dimer coverings on a square lattice of size $n\times n$ with periodic boundary conditions.
The bijection $O_{rs}$ can be written as follows, 
\begin{equation}
    \begin{aligned}
        O_{rs}: ~&\mathbb{S}_{n,T} \rightarrow \mathbb{S}_{n,T},\\
        &S_T:=(\vb{a}_0,\vb{a}_1,\dots,\vb{a}_{n-1}) \mapsto (\vb{a}_1,\dots,\vb{a}_{n-1},\vb{a}_0),
    \end{aligned}
    \label{eq:group_action_element}
\end{equation}
where $\vb{a}_0,\dots,\vb{a}_{n-1}$ are the $n$ columns with $n$ sites each of the square lattice. 

We can use the symmetry structure of the generators now to reduce the number of dimer coverings that we have to consider.
The generators of the symmetries form a group.
Their action can be faithfully represented by the action of the group elements as in Eq.~\eqref{eq:group_action_element}.
Each unique class of dimer configurations is given by one of the orbits of the group action applied to the set of dimer coverings.
Since the symmetries commute with the Hamiltonian, \epscrit is identical in each orbit.
Thus, it is sufficient to compute \epscrit for one representative from each orbit.
In the following, we will call the different orbits a \emph{class} of dimer coverings.
In Section~\ref{sec:critical_epsilon} the equivalence of different dimer coverings in the same class is used to benchmark the algorithm.
Further details on the group structure can be found in Appendix~\ref{app:topology}.

In practice, the group orbits are obtained by a graph-exploration algorithm through a pre-generated list of all maximal dimer coverings.
Here, we pick depth-first search (DFS) for ease of implementation.
This list is obtained with a backtracking procedure.
The idea of the backtracking algorithm is to explore all valid configurations of maximal dimer coverings in a structured way.
The algorithm places the dimers successively on the lattice while checking for contradictions with the dimer constraint.
If a contradiction is found, all further attempts are stopped (i.e. this branch of the recursive search is discarded) and the algorithm continues from the last valid configuration with another placement strategy.

Given the list of maximal dimer coverings, we can classify them.
Starting from the first dimer covering, we recursively apply the generators of the symmetry group.
At each level of the recursion, we first apply a new operation before exploring other paths (following the depth-first strategy).
By tracking the already visited configurations, we discover a given orbit of the group action since we cannot leave the starting configuration's orbit by applying only symmetry operations.
After exploring the full orbit, we pick a new dimer covering that has not been visited before.
It must belong to a different orbit (class).
The procedure ends when all dimer coverings in the list have been visited.
For further details on the algorithm, see Appendix~\ref{app:dfs}.

\begin{figure}
    \centering
    \includegraphics[width=\columnwidth]{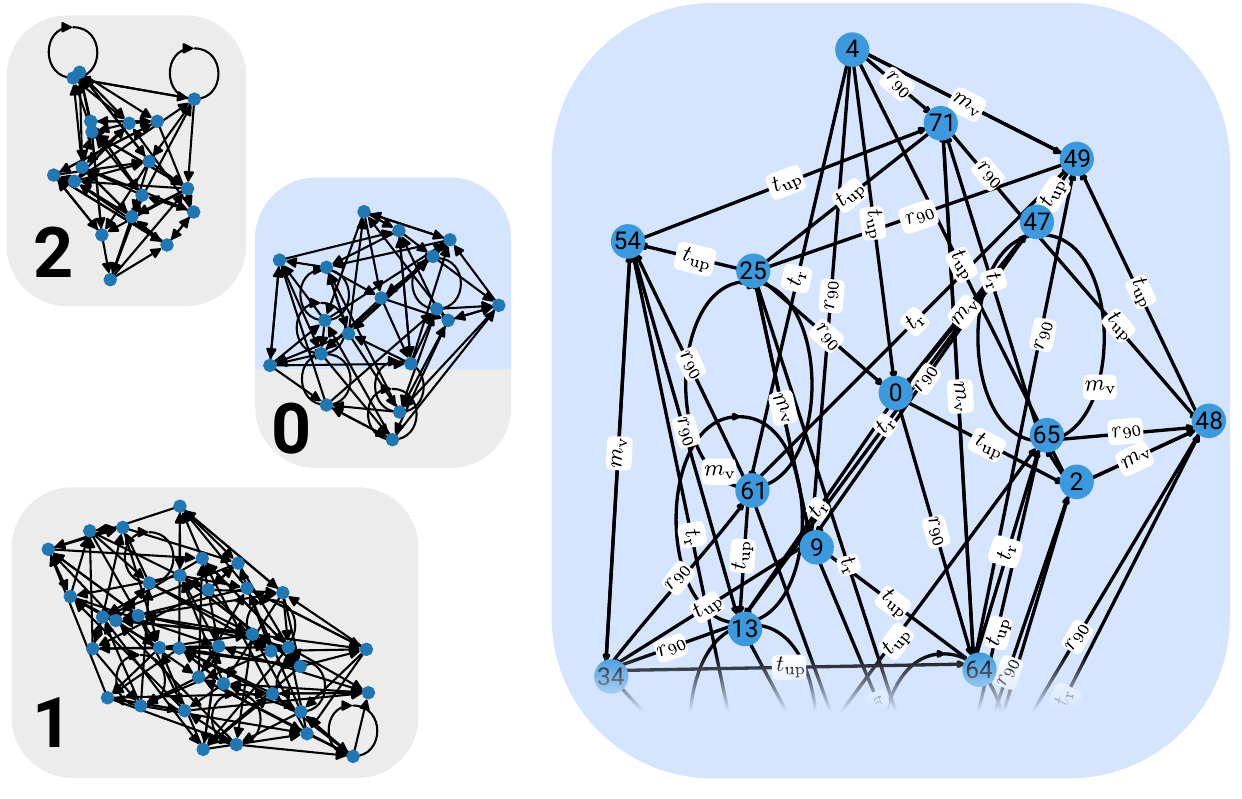}
    \caption{Graph orbits of a $3\times 3$ torus. 
    The dimers form three distinct classes, visible as three connected subgraphs.
    Each vertex corresponds to one dimer covering.
    The labels on the edges represent the symmetry operations that connect the different coverings.
    The shaded region in blue is a zoomed-in version of class $0$.}
    \label{fig:graph_orbits}
\end{figure}

As an example, we consider a square lattice of $3\times 3$ sites on the torus. Figure~\ref{fig:graph_orbits} shows the three classes of dimer coverings as discovered with DFS.
Each vertex of the graph corresponds to one maximal dimer covering.
They are connected by directed edges labeled with the operation that connects the two.
Instead of computing $\epscrit$ for all 72 dimer coverings, we can compute the value for only three classes.

An additional benefit of the DFS is an implicit check of the backtracking procedure.
The backtracking procedure places dimers on the lattice without any awareness of the symmetries.
The DFS applies all symmetry generators to all known dimer coverings.
During the DFS we only find known dimer coverings during the DFS algorithm; a necessary condition for the correctness of the backtracking algorithm.

\subsection{Computing the classical bound~\label{sec:classical_bound}}
After reducing the number of dimer coverings, we turn our attention to the computation of $\epscrit$, cf. Eq.~\eqref{eq:def_eps_crit}.
We start by computing the classical bound \clbound.
Its computation is equivalent to finding the optimal set of LDS for each party.
This assignment of local strategies minimizes the classical bound \clbound of the Bell inequality.

The number of LDS in a system of $N$ parties grows exponentially with $N$ as $(md)^N$.
Thus, solving the problem by fully enumerating all possible combinations quickly becomes prohibitively expensive.
For small systems, e.g. nine sites, the number of LDSs is $4^9=262144$ which is still manageable.
Already at system sizes of $4\times 4$ a more sophisticated approach is needed.

\subsubsection{Tropical Tensor Networks}
Tropical tensor networks are a more efficient way to obtain results for discrete optimization problems, e.g. the ground state of classical spin systems~\cite{liu_tropical_2021} or classical bounds of Bell inequalities~\cite{hu_tropical_2022}.
As the name suggests, we need to main ingredients: tropical algebra~\cite{maclagan_introduction_2021} and tensor networks~\cite{cirac_matrix_2021}.

Tropical algebra is defined on the tropical semiring $(\mathbb {R} \cup \{+\infty \},\oplus ,\odot)$, where the tropical addition $\oplus$ and tropical multiplication $\odot$ are defined as 
\begin{align}
    x\oplus y=\min\{x,y\},\quad   
    x\odot y=x+y.
\end{align}
This min-plus algebra yields a natural framework to formulate optimization problems~\cite{ebadi_quantum_2022}, e.g. the optimization of the classical bound \clbound.
A typical example of a graph optimization problem is given in Appendix~\ref{app:tropical}.

In the framework of tropical tensor networks, it is possible to interpret functions with finite number of possible inputs as tensors.
The functions $f(s_i,s_{i+1})$ describing the Bell inequality in the classical bound accepts only $d$ discrete inputs, corresponding to the number of local deterministic strategies.
To simplify, let us start by restricting to 1D lattices. 
In particular, let us denote the $N$ nodes of a chain as $i \in [N]$ and the strategy of each node is $s_i$.

\begin{figure}
    \includegraphics[width=\columnwidth]{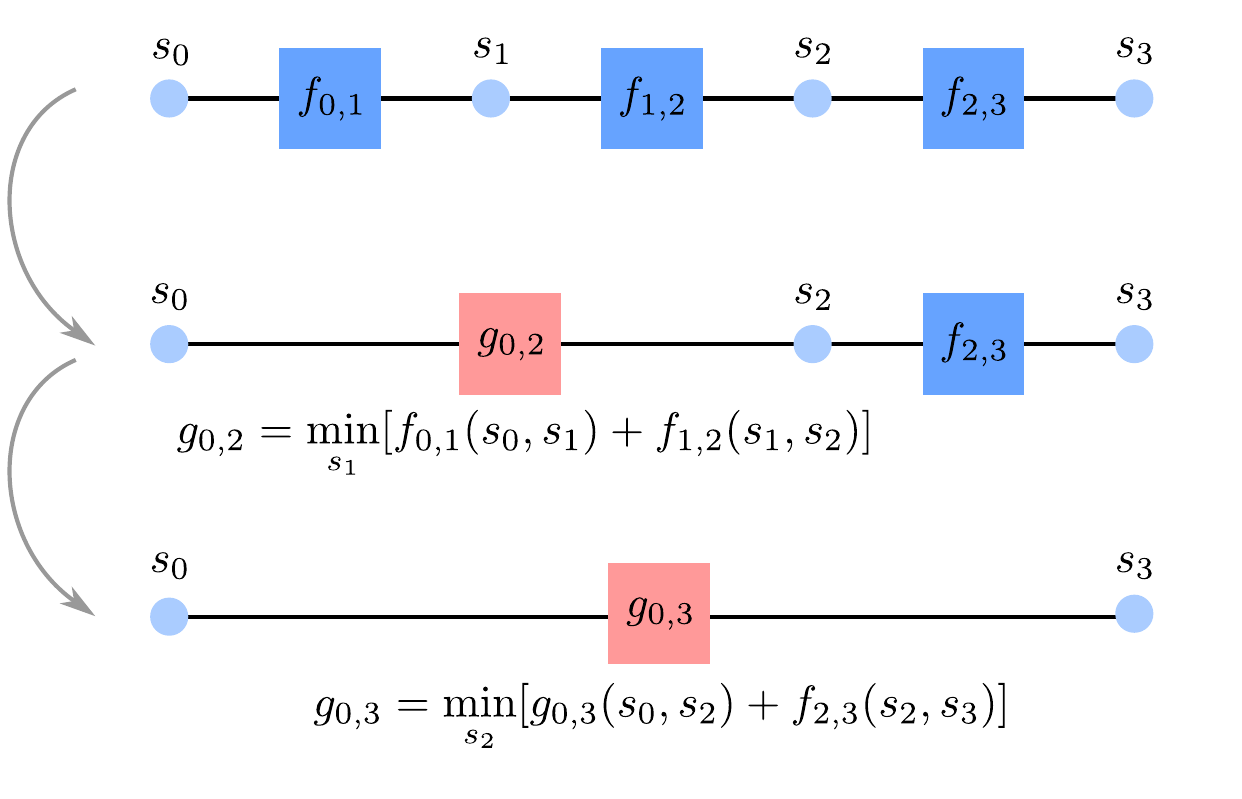}
    \caption{Example of successive contractions in one dimension. 
        In each step, one variable is eliminated by minimizing over it.
    }
    \label{fig:dynamic_programming_1d}
\end{figure}

Then as defined in Eq.~\eqref{eq:lds_Bell_ineq}, a Bell inequality involving at most nearest neighbour interactions will be a linear combination of functions $f_{i,i+1}:=f(s_i,s_{i+1})$ (see Figure~\ref{fig:dynamic_programming_1d}) such that
\begin{equation}
    \sum\limits_{i=0}^{\tilde{N}} f(s_i,s_{i+1}) - \clbound \geq 0,
\end{equation}
and the classical bound is 
\begin{equation}
   \clbound:=\min\limits_{\bf{s}\in \LDSset^{\tilde{N}}}\sum\limits_{i=0}^{\tilde{N}} f(s_i,s_{i+1})   
   \label{eq:def_classical_bound}
\end{equation}
Notice that we use $\tilde{N}=N-2$ for open boundary conditions or $\tilde{N}=N-1$ for periodic boundary conditions respectively.

The form of Eq.~\eqref{eq:def_classical_bound} is a formulation amenable to tropical tensor network contractions, significantly increasing the performance to obtain \clbound.
The idea here is to optimize one strategy $s_i$ at a time instead of all strategies $\bf{s}$ at once. 
This can be achieved by introducing a function 
\begin{equation}
    g_{i,i+2}:=\min\limits_{s_{i+1}}\left(f_{i,i+1}(s_{i},s_{i+1})+f_{i+1,i+2}(s_{i+1},s_{i+2})\right)
    \label{eq:elimination_third_party_1d}
\end{equation}
which optimizes over the strategy $s_{i+1}$ and effectively removes it as illustrated in Figure~\ref{fig:dynamic_programming_1d}. 
By iterating the function $g_{i,i+2}$ for the remaining nodes, a situation with only two remaining nodes is reached. 
At this point, one can efficiently obtain the final optimal value \clbound.

To complete the mapping from function minimization to tensor networks, we express the functions over discrete sets as tensors.
In the case of CHSH $d=2$ and $f$ can be fully described by a $4\times 4$ matrix $F$.
Thus, the minimization over a party (cf. Eq.~\eqref{eq:elimination_third_party_1d}) can be written as 
\begin{align}
    G=F\odot F,
\end{align}
where the matrix $G$ corresponds to the new function $g(s_i,s_{i+2})$ in Eq.~\eqref{eq:elimination_third_party_1d} and $\odot$ stands for tropical matrix multiplication.
Adhering to standard tensor network notation, we can write the full dynamic programming approach including the iteration over all sites as a contraction
\begin{align}
    \begin{aligned}
       \clbound&=\min\limits_{\bf{s}\in \LDSset^N}\sum\limits_{i=0}^{N-1} f(s_i,s_{i+1})\\
       &=\vcenter{\hbox{\includegraphics[width=0.6\columnwidth]{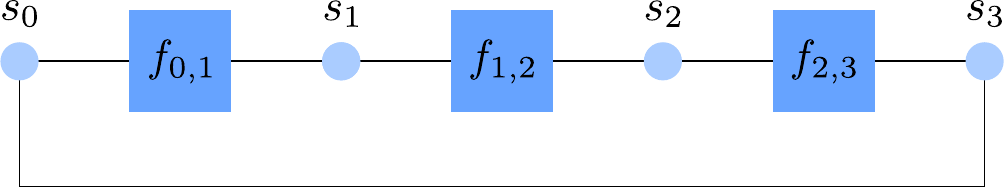}}}.
    \end{aligned}
\end{align}
Here, the tensors $s_i$ are delta distributions which are inserted for increased similarity with Figure~\ref{fig:dynamic_programming_1d}.
Leaving them out does not change the expression.
For further details on tropical tensor networks, we refer to Ref.~\cite{hu_tropical_2022}.

\subsubsection{Grouping by columns}
Let us now consider 2D lattices. 
The procedure to obtain \clbound follows the same guidelines presented in the previous section, but this time we are going to group the nodes by columns. 
The exact contraction of $2$D tensor networks scales exponentially with system size~\cite{schuch_computational_2007}.
In contrast to a one-dimensional lattice, the path during the iteration of $g_{i,i+2}$ is ambiguous.
Thus, we reduce the two-dimensional case to the one-dimensional case by grouping columns of the lattice.
This procedure exponentially increases the number of strategies per site.
It does not solve the issue of exponential scaling, but only confines it to one spatial direction.
In principle, we could investigate rectangular systems of limited height and large width.
Since we are considering square systems in this work to obtain a fair comparison between different sizes, the algorithm will only work for moderately sized lattices.

To group the columns, we label the nodes $(i,j)$ with strategy $s_{ij}$ and $i,j$ corresponding to the row and column respectively, this time we introduce a function $g_{\text{col }j,\text{col }j+1}:=g(\vb{s}_j,\vb{s}_{j+1})$, where $\vb{s}_j$ is the tuple of all the variables $s_{i,j}$ in the $j$-th column. 
Now in each contraction step we are going to implicitly optimize over all the nodes of one column in the following manner:
\begin{align}
    \label{eq:opt_1step}
    g_{\text{col }j, \text{col }j+2}= &\min_{\vb{s}_{j+1}}(\sum_i \underbrace{f(s_{i,j},s_{i,j+1}) + f(s_{i,j+1},s_{i,j+2})}_{\text{Crossed columns interactions}} \notag \\
    &+\sum_{k=0}^2\underbrace{\sum_i f(s_{i,j+k},s_{i+1,j+k})}_{\text{Interactions sharing column}}),
\end{align}
here $\vb{s}_{j+1}=s_{0,j+1},\dots ,s_{i,j+1},\dots $ are all the strategies of nodes in the $j+1$ column.

\begin{figure}
    \includegraphics[width=\columnwidth]{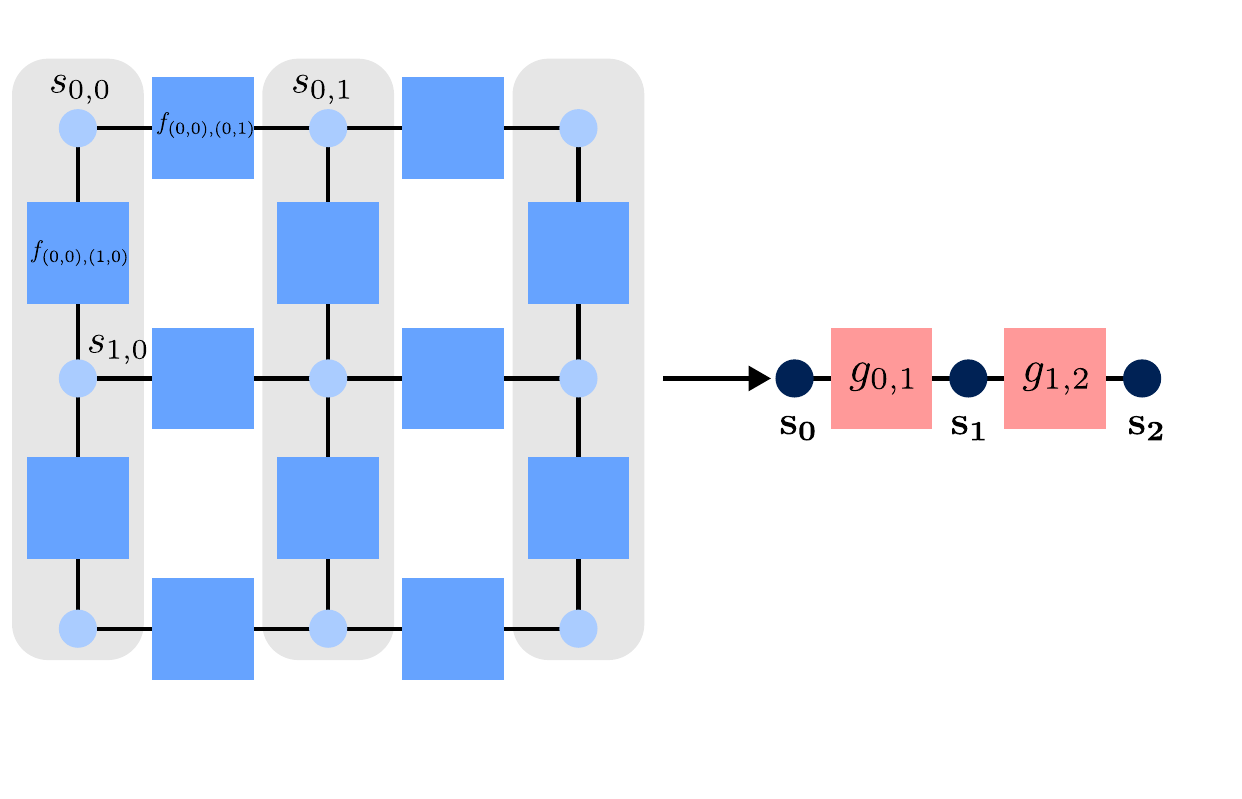}
    \caption{
        The two-dimensional system can be transformed into a one-dimensional system by blocking the columns to enlarged sites.
        The dimension of the variables $\vb{s}_i$ is exponentially bigger than the original variables $s_i$.
    }
    \label{fig:dynamic_programming_2d}
\end{figure}

Let us take Figure~\ref{fig:dynamic_programming_2d} for an explicit example.
In this case, one step in the contraction, for instance, would carry on the following optimization:
\begin{align}
\label{eq:example_opt_1step}
g_{0,2}=&\min_{s_{0,1},s_{1,1},s_{2,1}}\sum_{i=0}^2(f(s_{i,0},s_{i,1})+f(s_{i,1},s_{i,2}) \notag \\
&+\sum_{k=0}^2f(s_{i,0+k},s_{i+1,0+k})).
\end{align}
By grouping the sites in column $j$ into one variable $\vb{s}_j$, the optimization function $g_{0,2}$ then can be written as $g_{0,2}(\vb{s}_0,\vb{s}_2)=\min_{\vb{s}_1}g(\vb{s}_0,\vb{s}_1)+g(\vb{s}_1,\vb{s}_2).$

The approach of contracting tropical tensor networks can also be phrased in terms of dynamic programming~\cite{schuch_matrix_2010, aharonov_efficient_2010}.
The contracting in one spatial dimension is equivalent to the successive optimization steps in dynamic programming.

\subsection{Computing the quantum value\label{sec:quantum_bound}}
In addition to the classical bound \clbound, we need to compute \qbound, the quantum value of the Bell inequality.
Due to the structure of the Bell operator, the quantum value of the system corresponds to the ground state energy of $H$, the Hamiltonian associated to the Bell operator (cf. Section~\ref{sec:connection_ham}).
The problem of finding \qbound is equivalent to finding the ground state energy of $H$.
For small systems, we can obtain the ground state by diagonalization. 
For larger systems, however, this procedure becomes prohibitively expensive and we use dedicated many-body methods.

In contrast to the computation of the classical bound, we use matrix product states here as a computational tool with regular algebra.
Matrix product states (MPS) are one-dimensional tensor networks and we use them as an ansatz state in a variational optimization. 
Due to their entanglement structure, they target directly the ground state sector of local, gapped Hamiltonians in one space dimension~\cite{hastings_area_2007, arad_area_2013}.
Using variational methods like DMRG~\cite{white_density_1992}, they led to a deeper analytical and numerical understanding of many-body systems in one dimension~\cite{cirac_matrix_2021}. 
We aim to find a good approximation for the ground state energy by minimizing
\begin{align}
    E_{\textrm{min}}=\min_\alpha \frac{\bra{\psi(\alpha)}H\ket{\psi(\alpha)}}{\braket{\psi(\alpha)}},
\end{align}
where $\alpha$ is a set of matrices parameterizing the MPS.

Here, we apply MPS to two-dimensional systems by applying a snake-pattern~\cite{white_density_1998}.
This transforms the two-dimensional system into a one-dimensional system that we can optimize with DMRG.
This strategy introduces system-sized couplings in the Hamiltonian, limiting this approach to moderate system sizes.
For larger systems, genuine two-dimensional approaches like Projected Entangled Pair states (PEPS) would be more appropriate.
While the Hamiltonian obeys the boundary conditions demanded by the system, the MPS keeps open boundary conditions.
The boundary conditions are enforced by adding the couplings between the sides explicitly.
The open boundary conditions for the state are chosen due to the higher numerical efficiency.
Further details about the MPS simulations are given in Appendix~\ref{app:numerics}.

\subsection{Computing the critical epsilon\label{sec:critical_epsilon}}
In the last sections, we explored different methods to compute the classical bound \clbound and the quantum value \qbound.
Actually, we would like to compute the critical value of the coupling such that
\begin{align}
    \qbound(\epscrit)/\clbound(\epscrit)-1=0.
    \label{eq:eps_crit_root}
\end{align}
Due to the structure of the local polytope, we expect to find one critical value of epsilon on either side of $\epsilon=1$.

The root-finding procedure of Eq.~\eqref{eq:eps_crit_root} is performed by the iterative Brent-Dekker algorithm, a hybrid root-finding algorithm combining different root-finding methods~\cite{brent_chapter_1973, dekker_finding_1969}.
Given an initial value of $\epsilon$ on either side of $1$, the algorithm iteratively evaluates the numerator and denominator of Eq.~\eqref{eq:eps_crit_root} to find the critical value of $\epscrit$.
Since the quantum and the classical value are evaluated repeatedly, the parameters of both strategies have to be chosen with a time aspect in mind.
For more details on the numerical parameters, we refer to Appendix~\ref{app:numerics}.

\section{Results\label{sec:results}}
In the last section, we explored several methods to find classical bounds, quantum values and the critical value of $\epsilon$.
In the first step, we present benchmarks using exact methods like diagonalization for the variational simulations with tensor networks.
These benchmarks are presented in Section~\ref{sec:results_benchmarks}.
Our results for the critical value of $\epsilon$ for larger systems are shown in Section~\ref{sec:results_eps_crit}.

\subsection{Benchmark for small systems\label{sec:results_benchmarks}}
\begin{figure}
    \centering
    \includegraphics[width=\columnwidth]{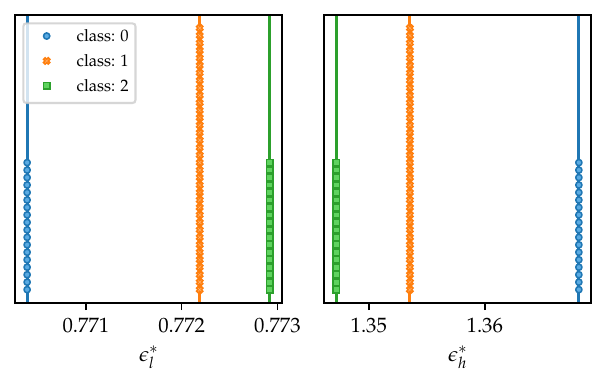}
    \caption{Comparison of exact results (vertical lines) and MPS results (dots) for $\epsilon^*$ on a torus of size $3\times 3$.
        The left(right) panel shows value of $\epsilon^*_l$($\epsilon^*_h$) smaller(greater) $1$.
        Each dot represents one dimer configuration on the lattice.
        The vertical axis only enumerates the different dimer configurations.}
    \label{fig:torus_L_03_ed_vs_mps}
\end{figure}

In small systems, both the Hilbert space and the total number of dimer coverings are still small.
Thus, the classical bound can be computed by enumerating all strategies and the quantum value can be evaluated by diagonalizing the Hamiltonian.
The vertical lines in Figure~\ref{fig:torus_L_03_ed_vs_mps} represent the result of these exact computations.
The illustration is split into two columns, one for each value $\epscrit$ (below and above $\epsilon=1$).
The dots represent $\epscrit$ for each dimer covering individually.
Here, the quantum value is calculated with MPS and the classical value results from a tropical tensor network (TrTN) contraction.
As expected, all points in Figure~\ref{fig:torus_L_03_ed_vs_mps} are located on the corresponding lines.
Thus, the computations converged to the expected values.
The plot serves as a benchmark for the variational computation since the values agree with large accuracy.

\subsection{Critical Epsilon\label{sec:results_eps_crit}}
After checking the convergence of the algorithm, we can compute \epscrit for larger systems.
The goal is to find combinations of a dimer configuration, system size, and boundary condition that allows to detect non-locality over large ranges of the coupling.
In the first step, we will explore the properties of the model and the violation ranges that it shows.
Afterwards, concrete dimer realizations with the maximal violation will be showcased.

\begin{figure*}
    \includegraphics[width=\textwidth]{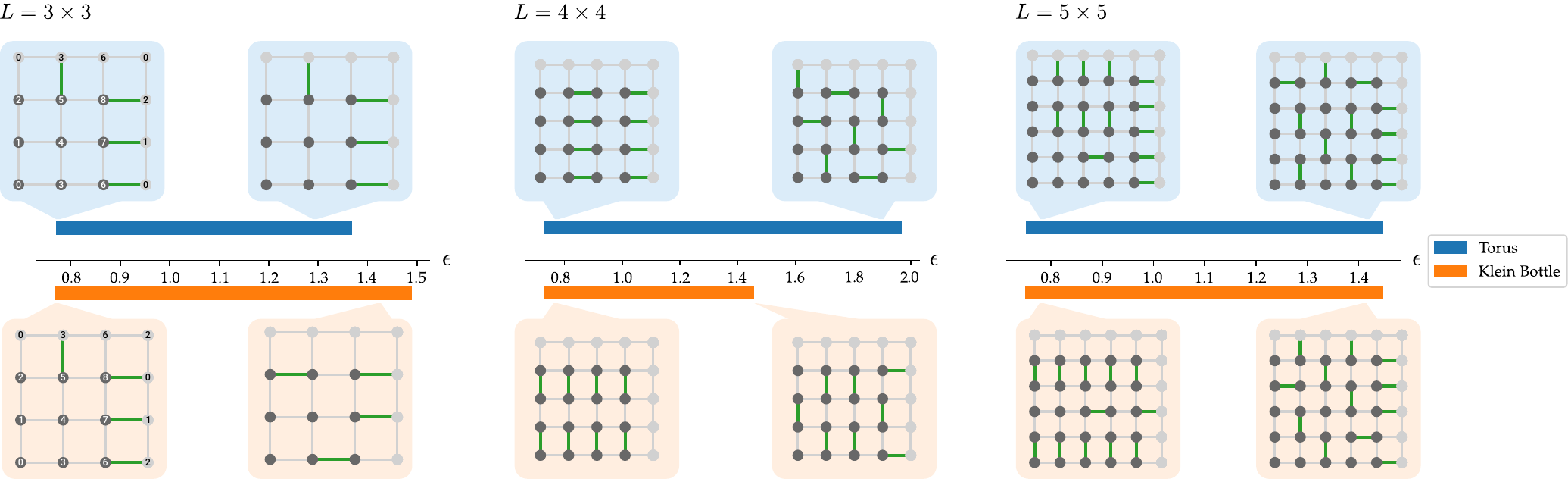}
    \caption{
        Ranges of critical values for different boundary conditions and system sizes.
        From left to right, the three figures indicate the results for $3\times 3, 4\times 4$ and $5\times 5$ systems.
        The blue(orange) regions shows the ranges of $\epscrit$ for a system on a torus(Klein bottle).
        The bars close to the axis connect the value of the minimal and the maximal critical epsilon for a given system. 
        The bars span across multiple classes.
        The insets indicate a dimer covering of the class with maximal(minimal) $\epscrit$.
    }
    \label{fig:overview_kb_torus_L_03-05}
\end{figure*}

In total, three different system sizes are investigated, square lattices of size $n\in\{3,4,5\}$.
Convergence for the first two lattice sizes were checked against exact diagonalization results.
Figure~\ref{fig:overview_kb_torus_L_03-05} shows the minimal $\epscrit_l$ and maximal $\epscrit_h$.
The size of the violation region depends on the chosen boundary conditions.
Furthermore, the size of the lattice plays a role.
With increasing lattice size, the difference between both the toroidal and Klein bottle boundaries decreases.
This could be expected since the bulk of the system grows faster than the boundary.
Since an extensive study of a $n=6$ lattice exceeded our numerical resources, it is not entirely clear whether it is an effect of the lattice size or the parity of the lattice dimensions.
The main problem is not the evaluation of a single model at a given coupling $\epsilon$, but rather the large amount of classes and the repeated evaluation during the root-finding process.

For all system sizes,  we observe a tendency to parallel ordering of the dimers for the lower bound of the violation interval.
On the upper end, perpendicular dimers are favored.
In the case of a system with $n=3$ on the Klein bottle, the same configuration realizes the lower and upper bound.
That makes this system a prime candidate for an experiment since the same dimer covering with different couplings will cover the whole interval.

\begin{figure}
    \centering
    \includegraphics[width=\columnwidth]{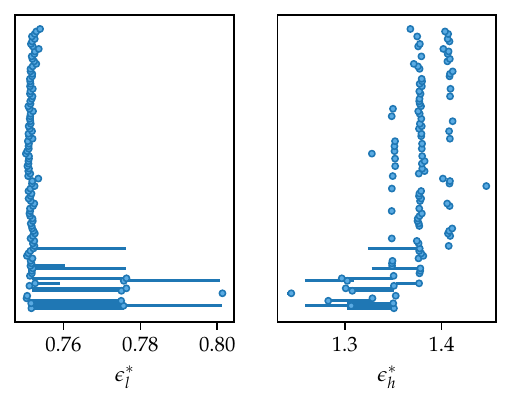}
    \caption{Computation of \epscrit for a $5\times 5$ lattice with periodic boundary conditions.
    The points are the median of 10 representative dimer configurations from each class.
    The asymmetric error bars show the minimal and maximal deviation among all considered realizations.}
    \label{fig:eps_crit_torus_L_05}
\end{figure}

For the two smaller system sizes, $n\in\{3,4\}$, all dimer configurations were evaluated to guarantee convergence of the root-finding procedure.
In the case of the larger lattice size, $n=5$, $10$ classes were evaluated for toroidal boundary conditions (see Fig.~\ref{fig:eps_crit_torus_L_05}).
Since the number of dimer configurations with Klein bottle boundary conditions exceeds $1000$, we only computed one representative dimer configuration for each class.

For systems of size $n=5$ and larger, the number of dimers exceeds the number of simulations that can be performed in a reasonable time.
Instead of simulating all dimers, we will actively use the dimer classification described in Section~\ref{sec:dimer_classification}.
Since the convergence of the MPS simulation becomes more challenging for larger systems, we simulate $10$ representatives from each class.
The results of the simulation are shown in Figure~\ref{fig:eps_crit_torus_L_05}.
Note that the vertical axis in Figure~\ref{fig:eps_crit_torus_L_05} does not display all dimer realizations as in Figure~\ref{fig:torus_L_03_ed_vs_mps}, but the different classes.
The points in the figure are the median values of the $10$ simulations with different dimer coverings belonging to the same class.
The asymmetric error bars represent the minimal and maximal value for the critical coupling among all runs for each class.
This error measure is more pessimistic than other error measures like the standard deviation.
We choose it here to take the highly asymmetric character of the error into account.
The main source of error is convergence accuracy in the MPS simulation.
Since this is a variational computation, the computation always overestimates $\qbound(\epsilon)$.
Depending on the slope of $\clbound(\epsilon)$ this leads either to an over- or an underestimation of \epscrit.

Finally, we can compare the ranges for the largest considered system size of $5\times 5$ (cf. right panel of Figure~\ref{fig:overview_kb_torus_L_03-05}).
In contrast to the smaller lattices, the difference between the torus and Klein bottle in terms becomes smaller with increasing lattice size.
One possible explanation is that the boundary effects should become less pronounced as the system grows.
The bulk of the system scales quicker than the boundary.
Due to run time considerations, the data for the Klein bottle is not averaged over 10 independent runs.
It represents the analysis of a randomly chosen dimer for each class.
Due to the large number of 1096 different dimer classes, we aimed to limit the computational time.

In an experimental setting, a configuration with as little fine-tuning of $\epsilon$ is desirable.
Dimer coverings with the largest violation range are most interesting, instead of the extremal cases for a given model.
In the considered systems, the configurations with maximal violation always coincide with the configuration on the upper bound of the $\epscrit$ range (cf. Fig.~\ref{fig:overview_kb_torus_L_03-05}).
These configurations are potential candidates for experimental realization.
Due to the large violation range, it is unnecessary to fine-tune the coupling.
The large gap between the classical and quantum value allows for an on-device energy-minimization with variational methods, like variational quantum eigensolvers~\cite{peruzzo_variational_2014,tilly_variational_2022}.

\section{Conclusions and Outlook\label{sec:conclusion}}
Despite the immense complexity of studying Bell nonlocality in many-body systems, the exploration of Bell nonlocality in terms of nearest-neighbor dimer Hamiltonians is an accessible avenue.

We find the violation regions for a host of two-dimensional CHSH inequalities by optimizing the coupling of a dimer Hamiltonian.
The intersections of the classical bound and the quantum value signify the boundaries of the violation interval.
The larger the interval, the larger the Bell inequality's capacity to indicate nonlocality.
Both the classical bound and the quantum value are evaluated numerically with tensor methods.
For the classical bound, we use tropical tensor networks, while the quantum value is evaluated as the ground state of a DMRG computation.

Dimer coverings with maximal violation region are interesting candidates for experimental realizations.
The considered system sizes are well within reach and the coupling scheme in terms of dimers are practically realizable.
Furthermore, the inequalities could be tailored to the quality of individual links of, for instance, a superconducting device.
Links with coupling problems could be given a low weight, while the rest of the lattice is still optimized for the best dimer covering.
The optimization of boundary conditions and coupling configuration (in terms of dimer covering) gives a practical approach to, with the same resources, better certify the nonlocality generation capabilities of existing quantum processors.

Looking ahead, there are several possible improvements to our approach.
The used numerical methods could be optimized by incorporating further symmetries of the Hamiltonian. 
This is possible for both exact diagonalization and MPS computations.
Since the system is two-dimensional, also projected entangled pair states (PEPS) are another option to compute at least a variationally constrained quantum value~\cite{verstraete_renormalization_2004,schuch_peps_2010}.
Due to the iterative procedure when finding the boundaries of the violating interval, it will be important to choose the algorithms for computing the quantum value and classical bound with their runtime in mind.

The investigation of the square lattice is a choice.
Since superconducting devices are often based on heavy-hexagon~\cite{kim_evidence_2023} or honeycomb lattices~\cite{xu_non-abelian_2024}, our approach could be naturally extended to non-square lattices, possibly with trimer interactions~\cite{giudice_trimer_2022}.

\acknowledgements
The authors acknowledge fruitful discussions with Weikang Li.
P.E. and J.T. acknowledge the support received by the Dutch National Growth Fund
(NGF), as part of the Quantum Delta NL programme. 
P.E. additionally acknowledges the support received through the NWO-Quantum Technology programme (Grant No. NGF.1623.23.006).
J.T. acknowledges the support received from the European Union’s Horizon Europe research and innovation programme through the ERC StG FINE-TEA-SQUAD (Grant No. 101040729). 
This publication is part of the ‘Quantum Inspire – the Dutch Quantum Computer in the Cloud’ project (with project number [NWA.1292.19.194]) of the NWA research program ‘Research on Routes by Consortia (ORC)’, which is funded by the Netherlands Organization for Scientific Research (NWO).
A.A. acknowledges support from the Austrian Science Fund (FWF) via project P 33730-N and by the ESQ Discovery programme (Erwin Schr{\"o}dinger Center for Quantum Science \& Technology), hosted by the Austrian Academy of Sciences ({\"O}AW). 
Parts of this work were performed by using the compute resources from the Academic Leiden Interdisciplinary Cluster Environment (ALICE) provided by Leiden University.
The views and opinions expressed here are solely those of the authors and do not necessarily reflect those of the funding institutions. 
Neither of the funding institutions can be held responsible for them.

\appendix
\section{Connection from Hamiltonians to Bell Inequalities\label{app:chained_bell}}
Here we show how to construct the multipartite Bell inequality with $m$ inputs and $2$ outcomes on a square $n\times n$ $2$D dimer coverings. 
Then conversely, for a given quantum spin Hamiltonian, we explain how to find the Bell inequality such that its Bell operator coincides with this Hamiltonian. 

To construct the multipartite Bell inequality of $m$-inputs and $2$-outcomes on square $n\times n$ $2$D lattices, 
we pick a dimer covering of the lattice first, then place a higher weight on the links within the dimer. 
Then given a dimer covering, one can construct its corresponding multipartite Bell inequality
\begin{align}
     \label{eq:multiBI2md_construct}
        I(\epsilon):=\sum_{\langle i,j\rangle}f_{i,j}(\epsilon) \cdot \mathcal{I}^{(i,j)}_{2,m,2},
\end{align}
where $f_{i,j}(\epsilon)$ is as defined in Eq.~\eqref{eq:fij(epsilon)}
and 
$\mathcal{I}^{(i,j)}_{2,m,2}$ denotes the bipartite Bell expression between nodes $i$ and $j$. 
One can see that for each link between node $i$ and $j$ of the dimer, we assign a $\mathcal{I}^{(i,j)}_{2,m,2}$  with higher weight $(1+\epsilon)\mathcal{I}^{(i,j)}$ associated to it. 
Similarly, we assign $(1-\epsilon)\mathcal{I}^{(i,j)}$ for the two adjacent nodes $i$ and $j$ that are not linked. 

Now, we illustrate the procedure for deriving Bell expressions corresponding to a given Hamiltonian of the form:
\begin{equation}
    \label{eq:givenH_general}
    H=\sum_{\langle i,j\rangle}f_{i,j}(\epsilon)H_2,
\end{equation}
where 
\begin{equation}
    \begin{aligned}
        H_2=&m\left(\cos^2 \frac{\pi}{2m}\sigma_x^{(i)}\sigma_x^{(j)}+\cos\frac{\pi}{2m}\sin\frac{\pi}{2m}\sigma_x^{(i)}\sigma_z^{(j)} \right.\\
        &\left.+\sin\frac{\pi}{2m}\cos\frac{\pi}{2m}\sigma_z^{(i)}\sigma_x^{(j)}-\cos^2\frac{\pi}{2m}\sigma_z^{(i)}\sigma_z^{(j)}\right), \notag
    \end{aligned}
\end{equation}
and 
$f_{i,j}(\epsilon)$ is defined as in Eq.~\eqref{eq:fij(epsilon)}. Note that this Hamiltonian $H$ is a particular case of the Bell inequality $I(\epsilon)$ in ~\eqref{eq:multiBI2md_construct} when $\mathcal{I}^{(i,j)}_{2,m,2}$ is the chained Bell inequality~\cite{braunstein_wringing_1990,tura_energy_2017}.
Our goal is to find a Bell operator $\mathcal{B}$ that corresponds to the given Hamiltonian $H$ in Eq.~\eqref{eq:givenH_general} such that $\mathcal{B} \equiv H$.
If we restrict to the local part of the Hamiltonian, then the structure of $H_2$ requires a specific Bell scenario: the number of parties in local parts is two because of the tensor form of $H_2$, and the number of outcomes $d =2$ due to the local dimension of the Pauli matrices. 
Thus for the local part $H_2$,
we only need to consider the Bell scenario $(2,m,2)$.  
To have non-trivial correlations, we set $m \geq 2$. 
According to the general form of Bell expression in $(2,m,2)$, the associated Bell operator can be written as 
\begin{align}
\mathcal{B}_2=\sum_{x_1,x_2=0}^{m-1}\sum_{k_1,k_2=0}^{1}\alpha_{x_1,x_2}^{(k_1,k_2)} A_{1,x_1}^{(k_1)}A_{2,x_2}^{(k_2)},
\end{align} 
where $A_{i,x_i}^{(k_i)}=\sum_{a_i=0}^{1}(-1)^{a_ik_i}F_{x_i,a_i}$ is the discrete Fourier transform of a positive operator-valued measure (POVM) $\{F_{x_i,a_i}\}_{a_i=0}^{1}$ representing the measurement on the $i$-th party in the basis $x_i$. Note that $[A_{1,x_1}^{(k_1)},A_{2,x_2}^{(k_2)}] = 0$ for $x_1,x_2 \in [m]$, $k_1,k_2 \in \{0,1\}$. 

Our goal is to find operators $A_{1,x_1}^{(k_1)},A_{2,x_2}^{(k_2)}$ and coefficients $\alpha_{x_1,x_2}^{(k_1,k_2)}$ that give rise to a non-trival Bell inequality. Due to the expression of the Hamiltonian $H$ in Eq.~\eqref{eq:givenH_general}, we assume that 
\begin{equation}
    \begin{aligned}
     \label{eq:operator}
    A_{1,x_1}^{(k_1)}&=\cos \theta_{x_1}^{(k_1)}\sigma_x+\sin\theta_{x_1}^{(k_1)}\sigma_z, \\
    A_{2,x_2}^{(k_2)}&=\cos \phi_{x_2}^{(k_2)}\sigma_x+\sin\phi_{x_2}^{(k_2)}\sigma_z,
    \end{aligned}
\end{equation}
where $x_1,x_2 \in [m]$, $k_1,k_2 \in \{0,1\}$. As we will see in the example below in ~\eqref{eq:eg_chsh_mapping}, by choosing $A_{1,x_1}^{(k_1)}$ and $A_{2,x_2}^{(k_2)}$ in this way, the matrix $T$ will have a desirable form.
Then the general form of the Bell operator can be written as 
\begin{align}
    \label{eq:belloperator2md}
    \mathcal{B}_2=&\sum_{x_1,x_2=0}^{m-1}\sum_{k_1,k_2=0}^{1}\alpha_{x_1,x_2}^{(k_1,k_2)}\cdot \left(\cos \theta_{x_1}^{(k_1)}\sigma_x+\sin\theta_{x_1}^{(k_1)}\sigma_z\right)  \notag \\
    &
    \otimes \left(\cos \phi_{x_2}^{(k_2)}\sigma_x+\sin\phi_{x_2}^{(k_2)}\sigma_z\right).
\end{align}
Next, to find the coefficients $\alpha_{x_1,x_2}^{(k_1,k_2)}$ of the Bell expression, we write the above Bell operator as a system of linear equations by projecting into the basis $\{\sigma_x\otimes \sigma_x,\sigma_x\otimes \sigma_z, \sigma_z\otimes \sigma_x,\sigma_z\otimes \sigma_z\}$. 
And the projection of $H_2$ is given by $ \operatorname{Tr}((\sigma_i\otimes \sigma_j)H_2)$, $i,j \in \{x,z\}$.
In this way, we have 
    \begin{align}
        T\cdot \vec{\alpha}=\vec{b},
    \end{align}
where 
\begin{equation}
    \begin{aligned}
         \label{eq:coefficientmatrix}
    T=	\begin{pmatrix}
         \cos \theta_{0}^{(0)}\cos \phi_{0}^{(0)}\quad \dots \quad	 \cos \theta_{x_1}^{(k_1)}\cos\phi_{x_2}^{(k_2)} \quad\dots 	\\
        \cos \theta_{0}^{(0)}\sin\phi_{0}^{(0)} \quad\dots   \quad \cos \theta_{x_1}^{(k_1)}\sin\phi_{x_2}^{(k_2)}  \quad\dots \\
             \sin \theta_{0}^{(0)}\cos\phi_{0}^{(0)} \quad\dots   \quad \sin \theta_{x_1}^{(k_1)}\cos\phi_{x_2}^{(k_2)}  \quad\dots  \\
             \sin \theta_{0}^{(0)}\sin \phi_{0}^{(0)} \quad \dots  \quad  \sin\theta_{x_1}^{(k_1)}\sin\phi_{x_2}^{(k_2)} \quad \dots 
        \end{pmatrix} \notag
    \end{aligned}
\end{equation}
is a $4\times 4m^2$ matrix, $ \vec{\alpha}=(\alpha_{0,0}^{0,0},\dots ,\alpha_{x_1,x_2}^{(k_1,k_2)},\dots ,\alpha_{m-1,m-1}^{(1,1)})^T$, and
    \begin{align}
        \vec{b}=&( \operatorname{Tr}((\sigma_x\otimes \sigma_x)H_2), \operatorname{Tr}((\sigma_x\otimes \sigma_z)H_2), \operatorname{Tr}((\sigma_z\otimes \sigma_x)H_2), \notag \\
        &\operatorname{Tr}((\sigma_z\otimes \sigma_z)H_2))^T, \notag \\
        =&2m(2\cos^2 \frac{\pi}{2m},\sin\frac{\pi}{m},\sin\frac{\pi}{m},-2\cos^2 \frac{\pi}{2m})^T.
    \end{align}
Assume that we fix $\theta_{x_1}^{(k_1)}, \phi_{x_2}^{(k_2)}$ for $x_1,x_2 \in [m]$, $k_1,k_2 \in \{0,1\}$, then we can write down the matrix $T$. Note that in our case, we need to consider the boundary conditions of the dimer coverings, so we assume $A_i=B_i$, which implies $\theta_{x_1}^{(k_1)}=\phi_{x_2}^{(k_2)}$. This ensures that the measurements of the site on the boundary are consistent.
Since $\rank(T)\leq \min\{4,4m^2\}$, we have the following two cases: If $T$ is invertible ($m=2$ and $\rank(T)=4$), there is a unique solution for $\vec{\alpha}=T^{-1}\vec{b}$. 
Thus there is a unique Bell expression corresponding to $H$ with operators given by Eq.~\eqref{eq:belloperator2md}. 
If $T$ is not invertible, there exists a family of solutions for $\vec{\alpha}$. 
It means there are multiple Bell expressions corresponding to $H$ with operators given by Eq.~\eqref{eq:belloperator2md}.

After obtaining the Bell expression $\mathcal{I}_{2,m,2}$ corresponding to the Bell operator in Eq.~\eqref{eq:belloperator2md}, one can construct the Bell expression corresponding to $H$ in Eq.~\eqref{eq:givenH_general} as follows,
    \begin{align}
            \label{eq:multiBI2md}
        I(\epsilon)=\sum_{\langle i,j\rangle}f_{i,j} \cdot \mathcal{I}^{(i,j)}_{2,m,2},
    \end{align}
where $f_{i,j}(\epsilon)$ is defined as in Eq.~\eqref{eq:fij(epsilon)} and $i,j$ label the parties.

Finally, as an illustrative example, we show how to find the associated operator of the CHSH inequality from the Hamiltonian in Eq.~\eqref{eq:givenH_general} when $m=2$. 
First,  
we assume that $\theta_0 = \phi_0=0 , \theta_1=\phi_1=\pi/2$,
then the operators are $A_0=B_0=\sigma_x$, $A_1=B_1=\sigma_z$. The Bell operator of the $(2,2,2)$ scenario without local operators is $\sum_{x,y=0,1}\alpha_{x,y} A_xB_y$, which  can be written as a system of linear equations by projecting into the basis $\{\sigma_i\otimes \sigma_j\}$, and the projection of
$H_2=(\sigma_x\sigma_x+\sigma_x\sigma_z+\sigma_z\sigma_x-\sigma_z \sigma_z)$ is given by $ \operatorname{Tr}((\sigma_i\otimes \sigma_j)H_2)$, $i,j \in \{x,z\}$. 
Then we have 
    \begin{align}
    \label{eq:eg_chsh_mapping}
        T\cdot \vec{\alpha}=\vec{b},
    \end{align}
where 
\begin{align}
    T=\begin{pmatrix}
        1 &0 & 0 & 0 \\
        0 & 1 & 0 & 0\\
        0 & 0 &1 & 0\\
        0 & 0& 0 & 1 
    \end{pmatrix}, \notag
\end{align}
and $ \vec{\alpha}=(\alpha_{0,0},\alpha_{0,1},\alpha_{1,0},\alpha_{1,1})^T$, $\vec{b}=(4,4,4,-4)^T$. Then one can obtain $\vec{\alpha}=T^{-1} \cdot \vec{b}=(4,4,4,-4)^T$.
The associated Bell expression is $\mathcal{I}^{(i,j)}=4(A_0B_0+A_0B_1+A_1B_0-A_1B_1)$. 
Finally, since $A_i,B_j\in \{-1,1\}$, $i,j \in\{0,1\}$, one obtains $\mathcal{I}^{(i,j)} \geq -8$, which is the (scaled) CHSH inequality. 
In this case, the Bell inequality associated with the Hamiltonian $H$ in Eq.~\eqref{eq:fromHtoBellineq} is 
\begin{align}
        \label{eq:multiBI222}
    I(\epsilon)=4\sum_{\langle i,j\rangle}f_{i,j}\mathcal{I}^{(i,j)},
\end{align}
where $\mathcal{I}^{(i,j)}=(A^{(i)}_0A^{(j)}_0+A^{(i)}_0A^{(j)}_1+A^{(i)}_1A^{(j)}_0-A^{(i)}_1A^{(j)}_1)$, $f_{i,j}(\epsilon)$ is defined as in Eq.~\eqref{eq:fij(epsilon)} and $i,j$ label the parties.

\section{Mathematical background for the classification of dimer configurations\label{app:topology}}
In two spatial dimension, the amount of possible dimer configurations increases dramatically with the number of sites.
However, some dimer configurations on the two-dimensional square lattice with fixed boundary conditions are related by symmetries.
This allows us to group them into a single class.
Subsequently, we only need to investigate a representative $2$D square dimer covering from each class, which allows us to reduce computational time.

Let $\mathbb{S}_n=\{S^{(1)},S^{(2)},\dots ,S^{(k)},\dots \}$ be a finite set of 2D square dimer coverings ($n \times n$ nodes) with boundary conditions, and let $G$ be a group with identity element $e$. 
Then a left action on $\mathbb{S}_n$ is a map $G\times \mathbb{S}_n \rightarrow \mathbb{S}_n$, written $(g,S^{(k)}) \mapsto g\cdot S^{(k)}$, such that 
\begin{eqnarray}
    g_1 \cdot (g_2 \cdot S^{(k)})=(g_1 \cdot g_2) \cdot S^{(k)}
\end{eqnarray} 
and $e\cdot S^{(k)}=S^{(k)}$ for all $g_1,g_2 \in G$ and $S^{(k)}  \in \mathbb{S}_n$. 

Let  $\mathbb{S}_{n,T}=\{S^{(1)}_T,S^{(2)}_T,\dots ,S^{(k)}_T,\dots \}$ be a finite set of 2D square lattice ($n \times n$ nodes) of torus (the boundary  $aba^{-1}b^{-1}$). 
For  $S^{(k)}_T \in \mathbb{S}_{n,T}$, we can obtain its equivalent dimer coverings with the same boundary conditions, $S^{(k)'}_T$, if some of the following operations are applied:
\begin{enumerate}
    \item Right shift $O_{rs}$. 
    The operation right shift is a bijection $O_{rs}$ from the set $\mathbb{S}_{n,T}$ to the set $\mathbb{S}_{n,T}$ as follows, 
    \begin{equation}
        \begin{aligned}
            &O_{rs}:\mathbb{S}_{n,T} \rightarrow \mathbb{S}_{n,T},\\
            &S^{(k)}_T:=(\vb{a}_0,\vb{a}_1,\dots,\vb{a}_{n-1}) \mapsto (\vb{a}_1,\dots,\vb{a}_{n-1},\vb{a}_0),  
        \end{aligned}
    \end{equation}
    where $\vb{a}_i$ is the $i$-th column with $n$ nodes of $S^{(k)}_T$, and $i=0,\dots ,n-1$.
    \item Up shift $O_{us}$.  
    The operation up shift $O_{us}$ is defined as 
    \begin{equation}
        \begin{aligned}
            &O_{us}:\mathbb{S}_{n,T} \rightarrow \mathbb{S}_{n,T},\\
            &S^{(k)}_T:=\begin{pmatrix}
                \vb{b}_0 \\
                \vdots\\
                \vb{b}_{n-2} \\
                \vb{b}_{n-1}
            \end{pmatrix} \mapsto \begin{pmatrix}
                \vb{b}_{n-1} \\
                \vb{b}_0 \\
                \vdots \\
                \vb{b}_{n-2}
            \end{pmatrix},
        \end{aligned}
    \end{equation}
    where $\vb{b}_j$ is the $j$-th row with $n$ nodes of $S^{(k)}_T$ and $j=0,\dots ,n-1$.
    \item Vertically mirrored $O_{vm}$. The operation vertical mirror is a bijection $O_{vm}$ from the set $S_{n,T}$ to the set $S_{n,T}$ as follows, 
    \begin{equation}
        \begin{aligned}
            & O_{vm}:\mathbb{S}_{n,T} \rightarrow \mathbb{S}_{n,T},\\
            & S^{(k)}_T:=(\vb{a}_0,\vb{a}_1,\dots ,\vb{a}_{n-1}) \mapsto (\vb{a}_{n-1},\dots ,\vb{a}_1,\vb{a}_0),
        \end{aligned}
    \end{equation}
    where $\vb{a}_i$ is the $i$-th column with $n$ nodes of $S^{(k)}_T$, and $i=0,\dots ,n-1$.
    \item Horizontally mirrored $O_{hm}$. The operation horizontal mirror $O_{hm}$ is 
    \begin{equation}
        \begin{aligned}
            &O_{hm}:\mathbb{S}_{n,T} \rightarrow \mathbb{S}_{n,T},\\
            & S^{(k)}_T:=\begin{pmatrix}
                \vb{b}_0 \\
                \vdots \\
                \vb{b}_{n-2} \\
                \vb{b}_{n-1}
            \end{pmatrix}
           \mapsto \begin{pmatrix}
               \vb{b}_{n-1} \\
               \vb{b}_{n-2} \\
               \vdots \\
               \vb{b}_0
           \end{pmatrix},
        \end{aligned}
    \end{equation}
    where $\vb{b}_j$ is the $j$-th row with $n$ nodes of $S^{(k)}_T$ and $j=0,\dots,n-1$.
    \item Rotation $O_{r}$. The operation rotation can be written as
    \begin{align}
        &O_{r}:\mathbb{S}_{n,T} \rightarrow \mathbb{S}_{n,T}, \\
        &S^{(k)}_T:=(\vb{a}_0,\vb{a}_1,\dots ,\vb{a}_{n-1})^T \mapsto \begin{pmatrix}
            \bar{\vb{a}}_{0}^T \\
            \bar{\vb{a}}_{1}^T \\
            \vdots \\
            \bar{\vb{a}}_{n-1}^T
        \end{pmatrix},
    \end{align}
    where $\vb{a}_i$ is the $i$-th column with $n$ nodes of $S^{(k)}_T$, and $\bar{\vb{a}}_{i}^T$ is the transpose after reversing the order of the elements along the length of the vector $\vb{a}_i$, and $i=0,\dots,n-1$. For example, if $\vb{a}_0=(2,1,0)^T$, then $\bar{\vb{a}}_0=(0,1,2)^T$ and $\bar{\vb{a}}_0^T=(0,1,2)$.
\end{enumerate}

\begin{figure}[h]
    \includegraphics[width=\columnwidth]{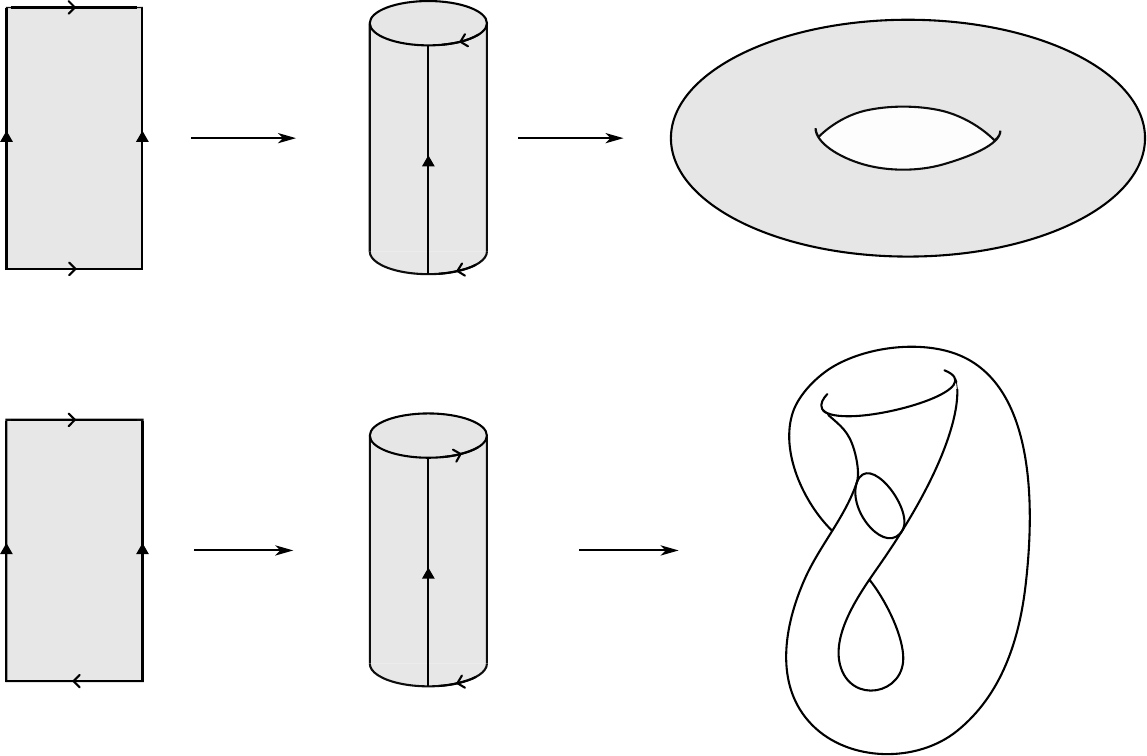}
    \caption{Sketch of the topology of a torus (top) and a Klein bottle (bottom). The arrows from left to right indicate the successive merging of boundaries.}
    \label{fig:topology}
\end{figure}   

One can check that for the nonempty set of 2D dimer coverings $\mathbb{S}_{n,T}$ of torus (the boundary  $aba^{-1}b^{-1}$) depicted in Fig.~\ref{fig:topology}, the group acting on $\mathbb{S}_{n,T}$ is 
\begin{equation}
    \begin{aligned}
        G_{T}=&\langle O_{rs},O_{us},O_{vm},O_{hm},O_{r}:\\&(O_{rs})^n=(O_{us})^n=(O_{vm})^2=(O_{hm})^2=(O_{r})^4=e,\\
        &O_{vm}O_{rs}=(O_{rs})^{-1}O_{vm},  O_{hm}O_{us}=(O_{us})^{-1}O_{hm},\\
        &O_{vm}O_{r}=(O_{r})^{-1}O_{vm},O_{hm}O_{r}=(O_{r})^{-1}O_{hm} \rangle.
    \end{aligned}
\end{equation}

Similarly, 	let  $\mathbb{S}_{n,KB}=\{S^{(1)}_{KB},S^{(2)}_{KB},\dots ,S^{(k)}_{KB},\dots \}$ be a finite set of 2D dimer coverings ($n \times n$ nodes) of Klein Bottle (the boundary $aba^{-1}b$) as shown in Fig. \ref{fig:topology}. For  $S^{(k)}_{KB} \in \mathbb{S}_{n,KB}$, we can obtain its equivalent dimer coverings with the same boundary conditions, $S'^{(k)}_{KB}$, if some of the following operations are applied:
\begin{enumerate}
    \item Right shift $O'_{rs}$. 
    The operation right shift is a bijection $O'_{rs}$ from the set $\mathbb{S}_{n,KB}$ to the set $\mathbb{S}_{n,KB}$ as follows, 
    \begin{equation}
        \begin{aligned}
            &O'_{rs}:\mathbb{S}_{n,KB} \rightarrow \mathbb{S}_{n,KB}, \\
            &S^{(k)}_{KB}:=(\vb{a}_0,\vb{a}_1,\dots ,\vb{a}_{n-1}) \mapsto (\vb{a}_1,\dots ,\vb{a}_{n-1},\bar{\vb{a}}_0),
        \end{aligned}
    \end{equation}
    where $\vb{a}_i$ is the $i$-th column with $n$ nodes of $S^{(k)}_{KB}$, $\bar{\vb{a}}_i$ reverses the order of the elements along the length of the vector $\vb{a}_i$ and $i=0,\dots ,n-1$.
    \item Vertically mirrored $O'_{vm}$. 
    The operation vertical mirror is a bijection $O'_{vm}$ from the set $\mathbb{S}_{n,KB}$ to the set $\mathbb{S}_{n,KB}$ as follows, 
    \begin{equation}
        \begin{aligned}
            &O'_{vm}:\mathbb{S}_{n,KB} \rightarrow \mathbb{S}_{n,KB}, \\
            &S^{(k)}_{KB}:=(\vb{a}_0,\vb{a}_1,\dots ,\vb{a}_{n-1}) \mapsto (\bar{\vb{a}}_0,\vb{a}_{n-1},\dots ,\vb{a}_1),
        \end{aligned}
    \end{equation}
    where $\vb{a}_i$ is the $i$-th column of $S^{(k)}_{KB}$ and $\bar{\vb{a}}_i$ reverses the order of the elements along the length of the vector $\vb{a}_i$, $i=0,\dots ,n-1$.
\end{enumerate} 
Then we can obtain that for the set $\mathbb{S}_{n,KB}$, the group acting on it is 
\begin{eqnarray}
    G_{KB}=\langle & O'_{rs},O'_{vm}:(O'_{rs})^{2n}(O'_{vm})^{2}=e, \notag \\
    &O'_{vm}O'_{rs}=(O'_{rs})^{-1}O'_{vm}\rangle. \notag \\
\end{eqnarray} 

\section{Dimer classification \label{app:dfs}}
Given the symmetries describe in Appendix~\ref{app:topology}, multiple dimer configurations lead to the same classical bounds and quantum values.
By considering only a single representative of each class, the amount of dimer configurations drops from 19600 individual dimers to 113 representatives.

As a classification procedure, we choose a depth first search.
Each dimer configuration is considered as a vertex of a graph.
Two vertices of the graph are connected by an edge if there exists a symmetry operation of the lattice transforming one dimer covering into the other.
To classify the dimers, we do not need to construct this graph explicitly, we only have to explore its connected components.
Each connected component corresponds to one distinct group orbit, i.e. a class of dimer coverings.
We start the classification procedure by choosing an arbitrary dimer configuration, i.e. an arbitrary vertex in the graph.
Following the spirit of a depth-first-search, we apply symmetry-operation of the lattice recursively to reach new vertices.
If a vertex has not been visited before, we mark it and apply a symmetry operation to the new vertex.
If it has been marked before, we perform no operation to exit the recursion.

Once the recursion terminates, all marked vertices belong to the same class.
We repeat the procedure until all vertices have been marked with a class label.

The number of classes varies depending on the different system sizes.
An overview of the amount of dimer classes as well as the minimal and maximal number of representatives in each class is given Table~\ref{tab:dimer_statistics}.
\begin{table}
    \centering
    \label{tab:dimer_statistics}
    \caption{Number of group orbits, i.e. different classes of dimer configurations.
    In parenthesis are given the minimal and the maximal number of dimer configurations in a class, respectively.}
    \begin{tabular}{ccc}
        \toprule
        \backslashbox{Size}{Boundary}& Torus  & Klein Bottle\\
        \midrule
        $3\times 3$ & 3 (18,36) & 11 (3,12)\\
        $4\times 4$ & 13 (4,64)& 36 (1,16)\\
        $5\times 5$ & 113 (50,200)& 1096 (5,20)\\
        \bottomrule
    \end{tabular}
\end{table}

\begin{figure}
    \centering
    \subfloat{
        \includegraphics[width=\columnwidth]{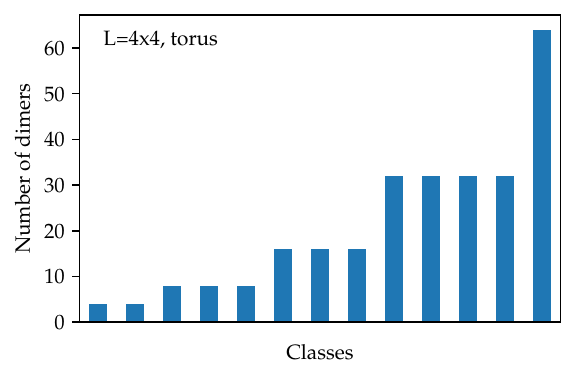}
    }\\
    \subfloat{
        \includegraphics[width=\columnwidth]{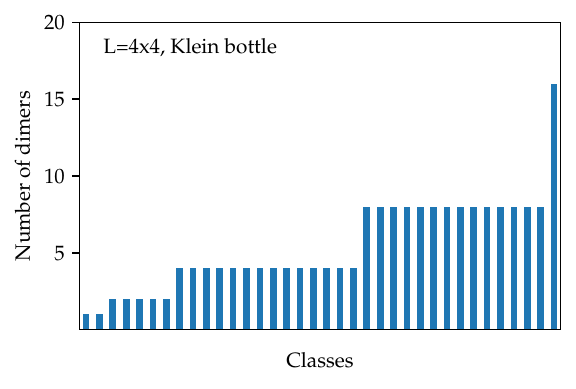}
    }
    \caption{Statistics of the dimer configurations. \emph{Top} Statistics for a system on a torus of size $4\times 4$. \emph{Bottom} Statistics for a system on a Klein bottle of the same size.}
    \label{fig:dimer_statistics}
\end{figure}

Figure~\ref{fig:dimer_statistics} shows the distribution of the dimer coverings over the different classes in more detail for periodic and Klein bottle boundary conditions in a system of size $4 \times 4$.
Since the numbering of the classes is arbitrary, the labels on the horizontal axis are left blank.
Here, the classes are ordered by number of dimer coverings for readability.

\section{Example Tropical Optimization\label{app:tropical}}
An example for an optimization problem is to find the shortest path in a directed graph in $k$ steps.
The directed graph $G=(V,E)$ is a tuple consisting of a set of vertices $V$ and a set of weighted, directed edges $E=(u,v,w)$, where $u,v\in V$ and $w\in\mathbb{R} \cup \{+\infty\}$ is the weight of the edge from vertex $u$ to $v$.
If there exists no edge from $u$ to $v$, we set $w=+\infty$. 

The graph can be equivalently represented by a $|V|\times|V|$ adjacency matrix $W$.
Each entry $W_{uv}$ of the matrix corresponds to the weight of the directed edge $(u,v)$.
This adjacency matrix is the input to the tropical optimization procedure.

The goal of the optimization is to find the shortest path in the graph from vertex $u$ to vertex $v$ in $k$ steps.
Here, \enquote{shortest} means the minimal amount of accumulated weight.
While this problem is a classical application for Dijkstra's algorithm, it can be formulated as tropical matrix multiplication.
The $(u,v)$ entry of the matrix $W^{\odot k}$ is the length of the shortest path in $k$ steps from vertex $u$ to $v$ in the directed graph $G$. 
Here, $W^{\odot k}$ is the tropical matrix power, i.e. applying tropical matrix multiplication $k$ times.
More concretely, we compute the tropical matrix product as
\begin{align}
    \begin{aligned}
        \left(A\odot B\right)_{ij}
        &=\bigoplus_{l} A_{il}\odot B_{lj}\\
        &=\min_l \left(A_{il} + B_{lj}\right).
    \end{aligned}
\end{align}
In the third line, the relation to a minimization task becomes evident.
The tropical matrix multiplication selects the minimum weight from all possible edges connecting the vertices $i$ and $j$.

As an example, let us consider the adjacency matrix 
\begin{align}
    W=\begin{pmatrix}
    1& 2 & +\infty \\
    +\infty & 3 & 4 \\
    5 &6 & 1 
    \end{pmatrix}.
\end{align}
To obtain the shortest path in two steps, we compute  
\begin{align}
    W^{\odot 2} =\begin{pmatrix}
    2 & 3& 6 \\
    9 & 6 & 5 \\
    6 & 7 & 2
    \end{pmatrix}.
\end{align}
The shortest path from the vertex $1$ to $2$ in two steps on graph $G$ is $3$, which corresponds to the entry $W^{\odot 2}_{1,2}$.

Additionally, the shortest closed path in $k$ steps can be obtained with the tropical trace, $\text{tropTrace}(W^{\odot k})$.
The tropical trace corresponds to taking the minimal diagonal entry of a matrix.
In our example, the minimum closed path in two steps is $2$, which is $\text{tropTrace}({W^{\odot 2}})$.

\section{Parameters for numerical simulations\label{app:numerics}}
The MPS simulations for the quantum were performed with TeNPy Library (version 0.10.0)~\cite{hauschild_efficient_2018}.
The MPS simulations are performed with a virtual bond dimension of $D=300$.
Since the ground state search is called repeteadly during the optimization, this bond dimension is a good compromise between accuracy and solution time.
The ground state is considered to be converged if the energy does not change more than $10^{-4}$ or the entropy does not change more than $10^{-3}$.

In the Brent-Dekker algorithm the root is considered to be converged if it does not change more than $10^{-3}$.

The code to generate the data used in this paper is available \href{https://gitlab.com/patrick.emonts/topo-bell/}{online}. 
The actual data used in this manuscript is available \href{https://gitlab.com/patrick.emonts/topo-bell-data}{here}.
\bibliography{references}
\end{document}